\documentclass [] {aa} 
\usepackage{epsfig, graphicx,longtable}
\usepackage{natbib}
\usepackage{times}

\bibpunct{(}{)}{;}{a}{}{,}

\begin{document}


\title{Spectroscopic [Fe/H] for 98 extra-solar planet-host stars\thanks{Based 
	       on observations collected at the La Silla Observatory, 
               ESO (Chile), with the CORALIE spectrograph 
               at the 1.2-m Euler Swiss telescope and the FEROS spectrograph
               at the 1.52-m and 2.2-m ESO telescopes, with the VLT/UT2 
               Kueyen telescope (Paranal Observatory, ESO, Chile) using the 
               UVES spectrograph (Observing run 67.C-0206, in service 
               mode), with the 
               TNG and William Herschel Telescopes, both operated at the 
               island of La Palma, and with the ELODIE spectrograph at the
               1.93-m telescope at the Observatoire de Haute Provence.}}
\subtitle{Exploring the probability of planet formation}

\author{
      N. C. Santos \inst{1,2}
 \and G. Israelian \inst{3}
 \and M. Mayor \inst{2}
}

\offprints{Nuno C. Santos, \email{Nuno.Santos@oal.ul.pt}}

\institute{
        Centro de Astronomia e Astrof{\'\i}sica da Universidade de Lisboa,
        Observat\'orio Astron\'omico de Lisboa, Tapada da Ajuda, 1349-018
        Lisboa, Portugal
     \and
	Observatoire de Gen\`eve, 51 ch.  des 
	Maillettes, CH--1290 Sauverny, Switzerland
     \and
	Instituto de Astrof{\'\i}sica de Canarias, E-38200 
        La Laguna, Tenerife, Spain}
\date{ACCEPTED FOR PUBLICATION IN ASTRONOMY\&ASTROPHYSICS } 

\titlerunning{Spectroscopic [Fe/H] for 98 extra-solar planet-host stars} 


\abstract{
We present stellar parameters and metallicities, 
obtained from a detailed spectroscopic analysis, for a large sample 
of 98 stars known to be orbited by planetary mass companions 
(almost all known targets), as well as for 
a volume-limited sample of 41 stars not known to host any planet.
For most of the stars the stellar parameters are revised versions
of the ones presented in our previous work. However, we also present
parameters for 18 stars with planets not previously published,
and a compilation of stellar parameters for the remaining 4 planet-hosts
for which we could not obtain a spectrum. 
A comparison of our stellar parameters with values of 
T$_{\mathrm{eff}}$, $\log{g}$, and [Fe/H] available in the literature
shows a remarkable agreement. In particular, our spectroscopic $\log{g}$
values are now very close to trigonometric $\log{g}$ estimates
based on Hipparcos parallaxes.
The derived [Fe/H] values are then used to confirm the previously 
known result that planets are more prevalent 
around metal-rich stars. Furthermore, we confirm that the frequency of 
planets is a strongly rising function
of the stellar metallicity, at least for stars with [Fe/H]$>$0. While only about 3\% of 
the solar metallicity stars in the CORALIE planet search sample
were found to be orbited by a planet, this number increases to more
than 25\% for stars with [Fe/H] above $+$0.3. Curiously, our results
also suggest that these percentages might
remain relatively constant for values of [Fe/H] lower than about 
solar, increasing then linearly with the mass fraction of heavy elements.
These results are discussed in the context of the theories of planetary
formation.
}

\maketitle

\section{Introduction}

The discovery of now more than 115 giant planets orbiting 
solar-type stars\footnote{See e.g. table at 
{\tt http://obswww.unige.ch/Exoplanets} for a continuously updated version} has lead to 
a number of different studies on the formation and evolution of the newly found planetary systems 
(for a recent review see e.g. \citet[][]{May03b} or \citet{San03b}). 
With the numbers increasing, current analyses are giving us the first 
statistically significant results about the properties of the new systems \citep[e.g.][]{Jor01,Zuc02,Udr03,San03,Egg03}.
Amongst these, some deal with the
planet-host stars themselves: they were found to be
significantly metal-rich with respect to the average field 
dwarfs \citep[e.g.][]{Gon97,Furetal98,Gon98,San00,San01,Gon01,Rei02,San03,Law03}.

Current studies seem to favor that this ``excess'' metallicity has
a primordial origin, i.e., that the high metal content of the
stars was common to the cloud of gas and dust that gave origin
to the star-planet system \citep[][]{Pin01,San01,San03}. Furthermore,
it has been shown that the frequency of planetary companions is
a strong function of the metal content of the 
star \citep[][]{San01,San03,Rei02,San03}: it is much easier to find planets around
metal-rich objects. Overall, the results
suggest that the formation of giant planets (or at least of
the kind we find now) is very dependent on the grain content of the
disk, a result that has important consequences for theories of
planetary formation \citep[][]{Pol96,Bos02,Ric03}.

During the last few years we have gathered spectra for planet host
stars, as well as of a sample of objects not known to harbor any
planetary companion. The main results of our uniform study, concerning the
metallicity of planet host stars, have been presented in \citet[][]{San00}, 
\citet[][]{San01}, and \citet[][]{San03} (hereafter Papers\,I, II, and III, respectively).

\begin{table*}
\caption[]{Atomic parameters and measured solar equivalent widths for 
the \ion{Fe}{i} and \ion{Fe}{ii} lines used.}
\begin{tabular}{lccc|lccc}
\hline
$\lambda$ (\AA) & $\chi_{l}$ & $\log{gf}$ & EW$_{\sun}$ (m\AA)& $\lambda$ (\AA) & $\chi_{l}$ & $\log{gf}$ & EW$_{\sun}$ (m\AA) \\
\hline
\ion{Fe}{i} & & & & 6591.32 & 4.59 & $-$1.98 & 10.6  \\ 
5044.22 & 2.85 & $-$2.04 & 73.4      & 6608.03 & 2.28 & $-$3.96 & 17.7  \\ 
5247.06 & 0.09 & $-$4.93 & 66.8      & 6627.55 & 4.55 & $-$1.48 & 28.0  \\ 
5322.05 & 2.28 & $-$2.90 & 60.4      & 6646.94 & 2.61 & $-$3.94 &  9.9  \\ 
5806.73 & 4.61 & $-$0.89 & 53.7      & 6653.86 & 4.15 & $-$2.41 & 10.5  \\ 
5852.22 & 4.55 & $-$1.19 & 40.6      & 6703.57 & 2.76 & $-$3.02 & 36.9  \\ 
5855.08 & 4.61 & $-$1.53 & 22.4      & 6710.32 & 1.48 & $-$4.82 & 16.0  \\ 
5856.09 & 4.29 & $-$1.56 & 33.8      & 6725.36 & 4.10 & $-$2.20 & 17.2  \\ 
6027.06 & 4.08 & $-$1.18 & 64.3      & 6726.67 & 4.61 & $-$1.05 & 46.9  \\ 
6056.01 & 4.73 & $-$0.50 & 72.4      & 6733.16 & 4.64 & $-$1.43 & 26.8  \\ 
6079.01 & 4.65 & $-$1.01 & 45.7      & 6750.16 & 2.42 & $-$2.61 & 74.1  \\ 
6089.57 & 5.02 & $-$0.88 & 35.0      & 6752.71 & 4.64 & $-$1.23 & 35.9  \\ 
6151.62 & 2.18 & $-$3.30 & 49.8      & 6786.86 & 4.19 & $-$1.90 & 25.2  \\ 
6157.73 & 4.07 & $-$1.24 & 61.9      &         &      &         &       \\
6159.38 & 4.61 & $-$1.86 & 12.4      & \ion{Fe}{ii} & &         &       \\
6165.36 & 4.14 & $-$1.50 & 44.6      & 5234.63 & 3.22 & $-$2.23 & 83.7  \\ 
6180.21 & 2.73 & $-$2.64 & 55.8      & 5991.38 & 3.15 & $-$3.53 & 31.5  \\ 
6188.00 & 3.94 & $-$1.63 & 47.7      & 6084.11 & 3.20 & $-$3.78 & 20.8  \\ 
6200.32 & 2.61 & $-$2.40 & 73.3      & 6149.25 & 3.89 & $-$2.72 & 36.2  \\ 
6226.74 & 3.88 & $-$2.07 & 29.3      & 6247.56 & 3.89 & $-$2.35 & 52.2  \\ 
6229.24 & 2.84 & $-$2.89 & 37.9      & 6369.46 & 2.89 & $-$4.13 & 19.2  \\ 
6240.65 & 2.22 & $-$3.29 & 48.3      & 6416.93 & 3.89 & $-$2.64 & 40.1  \\ 
6265.14 & 2.18 & $-$2.56 & 86.0      & 6432.69 & 2.89 & $-$3.56 & 41.5  \\ 
6270.23 & 2.86 & $-$2.58 & 52.3      & 6446.40 & 6.22 & $-$1.91 &  4.2  \\ 
6380.75 & 4.19 & $-$1.32 & 52.2      & 7479.70 & 3.89 & $-$3.59 & 10.0  \\ 
6392.54 & 2.28 & $-$3.93 & 18.1      & 7515.84 & 3.90 & $-$3.43 & 13.4  \\ 
6498.94 & 0.96 & $-$4.63 & 45.9      & 7711.73 & 3.90 & $-$2.55 & 46.0  \\ 
\hline
\end{tabular}
\label{tab1}
\end{table*}

Most groups working on exoplanets are now convinced 
that planet host stars are really
more metal-rich than average field dwarfs. This result is
clearly independent of the kind of analysis done to obtain the stellar
metallicity \citep[e.g.][]{Gim00,Gon01,San01,Mur02,Martell02,Hei03}, and in Paper\,III 
we showed that this
result is not due to any observational bias. However, some authors have
questioned the quality of the spectroscopic analyses we (and others) have been
publishing. In particular, the relatively high surface gravities derived in our
preceding papers led to some criticism regarding this matter.

In order to address this problem, in this paper we present a revised
spectroscopic analysis for all the stars presented in Papers\,II and III.
The new derived surface gravities are now compatible with the ones obtained 
by other authors, and with trigonometric gravities derived
using Hipparcos parallaxes \citep[][]{ESA97}. Other stellar 
parameters (T$_{\mathrm{eff}}$ and [Fe/H]) are also similar to 
the ones presented elsewhere in the literature, and not particularly
different from the ones derived in Papers\,II and III.

Furthermore, we have derived stellar parameters for 18 planet host stars not 
analyzed before, increasing to 98 the number of these
objects for which we have precise spectral information.
The new results unambiguously confirm the previously presented trends:
stars with planetary companions are more metal-rich than average 
field dwarfs.

\begin{table*}
\caption[]{Stars with planets and derived stellar parameters (HD number between 1 and 60\,000). See text for more details.}
\begin{tabular}{lcccrccccc}
\hline
HD     & T$_{\mathrm{eff}}$ & $\log{g}_{spec}$ & $\xi_{\mathrm{t}}$ & \multicolumn{1}{c}{[Fe/H]} & N(\ion{Fe}{i},\ion{Fe}{ii}) & $\sigma$(\ion{Fe}{i},\ion{Fe}{ii}) & Instr.$^a$ & Mass         & $\log{g}_{hipp}$ \\
number & [K]                &  [cm\,s$^{-2}$]  &  [km\,s$^{-1}$]  &        &                              &                                    &            & [M$_{\sun}$] & [cm\,s$^{-2}$] \\
\hline
\object{HD\,142   }    &  6302$\pm$56  &  4.34$\pm$0.13 &  1.86$\pm$0.17 &   0.14$\pm$0.07 & 28,8 & 0.05,0.05 & [2]  & 1.28 &4.27\\
\object{HD\,1237  }    &  5536$\pm$50  &  4.56$\pm$0.12 &  1.33$\pm$0.06 &   0.12$\pm$0.06 & 37,7 & 0.05,0.06 & [1]  & 0.99 &4.56\\
\object{HD\,2039  }    &  5976$\pm$51  &  4.45$\pm$0.10 &  1.26$\pm$0.07 &   0.32$\pm$0.06 & 34,6 & 0.05,0.04 & [1]  & 1.18 &4.35\\
\object{HD\,3651  }    &  5173$\pm$35  &  4.37$\pm$0.12 &  0.74$\pm$0.05 &   0.12$\pm$0.04 & 31,5 & 0.04,0.05 & [4]  & 0.76 &4.41\\
\object{HD\,4203  }    &  5636$\pm$40  &  4.23$\pm$0.14 &  1.12$\pm$0.05 &   0.40$\pm$0.05 & 37,7 & 0.05,0.07 & [2]  & 1.06 &4.19\\
\object{HD\,4208  }    &  5626$\pm$32  &  4.49$\pm$0.10 &  0.95$\pm$0.06 &$-$0.24$\pm$0.04 & 37,7 & 0.04,0.05 & [2]  & 0.86 &4.48\\
\object{HD\,6434  }    &  5835$\pm$59  &  4.60$\pm$0.12 &  1.53$\pm$0.27 &$-$0.52$\pm$0.08 & 30,4 & 0.06,0.06 & [2]  & 0.82 &4.33\\
\object{HD\,8574  }    &  6151$\pm$57  &  4.51$\pm$0.10 &  1.45$\pm$0.15 &   0.06$\pm$0.07 & 30,7 & 0.06,0.04 & [4]  & 1.18 &4.28\\
\object{HD\,9826  }    &  6212$\pm$64  &  4.26$\pm$0.13 &  1.69$\pm$0.16 &   0.13$\pm$0.08 & 27,6 & 0.06,0.05 & [4]  & 1.30 &4.16\\
\object{HD\,10647 }    &  6143$\pm$31  &  4.48$\pm$0.08 &  1.40$\pm$0.08 &$-$0.03$\pm$0.04 & 34,6 & 0.03,0.03 & [1]  & 1.14 &4.43\\
\object{HD\,10697 }    &  5641$\pm$28  &  4.05$\pm$0.05 &  1.13$\pm$0.03 &   0.14$\pm$0.04 & 33,7 & 0.03,0.03 & [4]  & 1.22 &4.03\\
\object{HD\,12661 }    &  5702$\pm$36  &  4.33$\pm$0.08 &  1.05$\pm$0.04 &   0.36$\pm$0.05 & 34,8 & 0.04,0.03 & [3]  & 1.05 &4.34\\
\object{HD\,13445 }    &  5119$\pm$43  &  4.48$\pm$0.14 &  0.63$\pm$0.07 &$-$0.25$\pm$0.05 & 38,6 & 0.05,0.07 & [1]  & 0.67 &4.44\\
\object{HD\,13445 }    &  5207$\pm$30  &  4.56$\pm$0.11 &  0.82$\pm$0.05 &$-$0.23$\pm$0.04 & 38,5 & 0.03,0.05 & [2]  & 0.74 &4.52\\
\object{HD\,13445 }    &  5163         &  4.52     	&  0.72          &\multicolumn{1}{c}{$-$0.24}     	   &      &           & avg. & 0.70 &4.48\\
\object{HD\,16141 }    &  5801$\pm$30  &  4.22$\pm$0.12 &  1.34$\pm$0.04 &   0.15$\pm$0.04 & 37,7 & 0.03,0.04 & [2]  & 1.05 &4.17\\
\object{HD\,17051 }    &  6252$\pm$53  &  4.61$\pm$0.16 &  1.18$\pm$0.10 &   0.26$\pm$0.06 & 34,6 & 0.05,0.07 & [2]  & 1.32 &4.49\\
\object{HD\,19994 }    &  6217$\pm$67  &  4.29$\pm$0.08 &  1.62$\pm$0.12 &   0.25$\pm$0.08 & 35,5 & 0.06,0.03 & [1]  & 1.37 &4.14\\
\object{HD\,19994 }    &  6290$\pm$58  &  4.31$\pm$0.13 &  1.63$\pm$0.12 &   0.32$\pm$0.07 & 33,6 & 0.06,0.05 & [2]  & 1.40 &4.17\\
\object{HD\,19994 }    &  6121$\pm$33  &  4.06$\pm$0.05 &  1.55$\pm$0.06 &   0.19$\pm$0.05 & 37,5 & 0.04,0.03 & [5]  & 1.34 &4.09\\
\object{HD\,19994 }    &  6132$\pm$67  &  4.11$\pm$0.23 &  1.37$\pm$0.12 &   0.21$\pm$0.08 & 35,6 & 0.06,0.09 & [3]  & 1.36 &4.10\\
\object{HD\,19994 }    &  6190         &  4.19     	&  1.54          &   \multicolumn{1}{c}{0.24}     	   &      &           & avg. & 1.37 &4.12\\
\object{HD\,20367 }    &  6138$\pm$79  &  4.53$\pm$0.22 &  1.22$\pm$0.16 &   0.17$\pm$0.10 & 31,6 & 0.08,0.09 & [6]  & 1.21 &4.42\\
\object{HD\,22049 }    &  5073$\pm$42  &  4.43$\pm$0.08 &  1.05$\pm$0.06 &$-$0.13$\pm$0.04 & 37,6 & 0.05,0.04 & [1]  & 0.73 &4.55\\
\object{HD\,23079 }    &  5959$\pm$46  &  4.35$\pm$0.12 &  1.20$\pm$0.10 &$-$0.11$\pm$0.06 & 35,6 & 0.05,0.05 & [2]  & 1.01 &4.36\\
\object{HD\,23596 }    &  6108$\pm$36  &  4.25$\pm$0.10 &  1.30$\pm$0.05 &   0.31$\pm$0.05 & 36,6 & 0.04,0.04 & [3]  & 1.30 &4.22\\
\object{HD\,27442 }    &  4825$\pm$107 &  3.55$\pm$0.32 &  1.18$\pm$0.12 &   0.39$\pm$0.13 & 36,6 & 0.11,0.13 & [2]  &  --  &--  \\
\object{HD\,28185 }    &  5656$\pm$44  &  4.45$\pm$0.08 &  1.01$\pm$0.06 &   0.22$\pm$0.05 & 38,6 & 0.05,0.03 & [1]  & 0.98 &4.39\\
\object{HD\,30177 }    &  5591$\pm$50  &  4.35$\pm$0.12 &  1.03$\pm$0.06 &   0.39$\pm$0.06 & 37,4 & 0.06,0.05 & [1]  & 1.01 &4.34\\
\object{HD\,30177 }    &  5584$\pm$65  &  4.23$\pm$0.13 &  1.14$\pm$0.07 &   0.38$\pm$0.09 & 38,7 & 0.07,0.05 & [2]  & 1.01 &4.34\\
\object{HD\,30177 }    &  5588         &  4.29     	&  1.08          &   \multicolumn{1}{c}{0.39}     	   &      &           & avg. & 1.01 &4.34\\
\object{HD\,33636$^b$} &  6046$\pm$49  &  4.71$\pm$0.09 &  1.79$\pm$0.19 &$-$0.08$\pm$0.06 & 37,6 & 0.05,0.04 & [2]  & 1.16 &4.56\\
\object{HD\,37124 }    &  5546$\pm$30  &  4.50$\pm$0.03 &  0.80$\pm$0.07 &$-$0.38$\pm$0.04 & 36,7 & 0.04,0.02 & [3]  & 0.75 &4.33\\
\object{HD\,38529 }    &  5674$\pm$40  &  3.94$\pm$0.12 &  1.38$\pm$0.05 &   0.40$\pm$0.06 & 34,7 & 0.05,0.06 & [2]  & 1.60 &3.81\\
\object{HD\,39091$^b$} &  5991$\pm$27  &  4.42$\pm$0.10 &  1.24$\pm$0.04 &   0.10$\pm$0.04 & 38,7 & 0.03,0.04 & [1]  & 1.10 &4.38\\
\object{HD\,40979 }    &  6145$\pm$42  &  4.31$\pm$0.15 &  1.29$\pm$0.09 &   0.21$\pm$0.05 & 24,9 & 0.04,0.07 & [4]  & 1.21 &4.38\\
\object{HD\,46375 }    &  5268$\pm$55  &  4.41$\pm$0.16 &  0.97$\pm$0.06 &   0.20$\pm$0.06 & 37,4 & 0.05,0.07 & [3]  & 0.82 &4.34\\
\object{HD\,47536 }    &  4554$\pm$85  &  2.48$\pm$0.23 &  1.82$\pm$0.08 &$-$0.54$\pm$0.12 & 37,6 & 0.11,0.09 & [2]  &   -- &--  \\
\object{HD\,49674 }    &  5644$\pm$54  &  4.37$\pm$0.07 &  0.89$\pm$0.07 &   0.33$\pm$0.06 & 33,5 & 0.06,0.04 & [4]  & 1.04 &4.50\\
\object{HD\,50554 }    &  6026$\pm$30  &  4.41$\pm$0.13 &  1.11$\pm$0.06 &   0.01$\pm$0.04 & 37,6 & 0.03,0.05 & [3]  & 1.09 &4.40\\
\object{HD\,52265 }    &  6076$\pm$57  &  4.20$\pm$0.17 &  1.38$\pm$0.09 &   0.20$\pm$0.07 & 39,7 & 0.06,0.07 & [1]  & 1.19 &4.32\\
\object{HD\,52265 }    &  6131$\pm$47  &  4.35$\pm$0.13 &  1.33$\pm$0.08 &   0.25$\pm$0.06 & 36,6 & 0.05,0.04 & [2]  & 1.21 &4.34\\
\object{HD\,52265 }    &  6103         &  4.28     	&  1.36          &   \multicolumn{1}{c}{0.23}     	   &      &           & avg. & 1.20 &4.33\\
\hline
\end{tabular}
\\ $^a$ The instruments used to obtain the spectra were: [1] 1.2-m Swiss Telescope/CORALIE; [2] 1.5-m and 2.2-m ESO/FEROS; [3] WHT/UES; [4] TNG/SARG; [5] VLT-UT2/UVES; [6] 1.93-m OHP/ELODIE; [7] Keck/HIRES
\newline
$^b$ The companions to these stars have minimum masses above 10\,M$_{\mathrm{Jup}}$, and are thus probably brown-dwarfs.  
\label{tabplan1}
\end{table*}

\section{The data}

Most of the spectra for the planet-host stars analyzed in this paper 
were studied in Papers I, II, and III. We refer the reader to these for a description of 
the data. 

{
During the last year, however, we have obtained spectra
for 18 more planet host stars. Most of the spectra were gathered using the 
FEROS spectrograph (2.2-m ESO/MPI telescope, La Silla, Chile),
on the night of the 12-13 March 2003 (for \object{HD\,47536}, \object{HD\,65216}, \object{HD\,72659}, \object{HD\,73256}, \object{HD\,73526}, \object{HD\,76700}, \object{HD\,111232}, and \object{HD\,142415}) and with the SARG spectrograph at the TNG telescope (La Palma, Spain) 
on the nights of the 9-10 October 2003 (for \object{HD\,3651}, \object{HD\,40979}, \object{HD\,68988}, \object{HD\,216770}, \object{HD\,219542B}, and \object{HD\,222404}). In these runs we 
have also gathered spectra for \object{HD\,30177}, \object{HD\,162020} (FEROS), and \object{HD178911B} (SARG), already previously analyzed. 
The FEROS spectra have S/N ratios above 300 for all targets at a resolution of 
about 50\,000, and were reduced using the FEROS pipeline software. The SARG spectra
have a resolution of about 57\,000, and were reduced using the tasks within the IRAF {\tt echelle} 
package\footnote{IRAF is distributed by National Optical Astronomy Observatories, operated 
by the Association of Universities for Research in Astronomy, Inc., 
under contract with the National Science Foundation, U.S.A.}.
Finally, a spectrum of \object{HD\,70642} with a S/N$\sim$150 was obtained using the CORALIE
spectrograph (R=50\,000), at the 1.2-m Euler Swiss telescope (La Silla, Chile),
on the night of the 21-22 October 2003. 

Equivalent Widths (EW) were measured using a Gaussian fitting procedure within the 
IRAF {\tt splot} task. For \object{HD\,178911\,B}, we also used the EW
measured by \citet[][]{Zuc01} from a Keck/HIRES spectrum (S. Zucker \& D. Latham, private communication). 
Given that only 16 \ion{Fe}{i} and 2 \ion{Fe}{ii} lines were measured from this spectrum,
the parameters derived are only listed as a test of consistency, but are not
used in rest of the paper.
Other previously obtained, but not used, spectra (see Paper\,III for the instrument 
description) were also analyzed for \object{HD\,89744} and \object{HD\,19994} (WHT/UES), 
\object{HD\,120136} (VLT/UVES), \object{HD\,49674} (TNG/SARG).

}

Besides the planet host stars, we also re-analyzed here our comparison sample of stars not 
known to harbor any planetary companion. This volume-limited sample, that represents a sub-sample of
the CORALIE planet search program stars \citep[][]{Udr00}, is described in Paper\,II. Since 2001, however, 
2 of the stars in the original list have been found to harbor planetary-mass companions: 
\object{HD\,39091} \citep[][]{Jon02} and \object{HD\,10647} \citep[][]{May03}. These are thus considered now
as planet hosts, adding to \object{HD1237}, \object{HD\,13445}, \object{HD\,17051}, 
\object{HD\,22049}, \object{HD\,217107}, also belonging to our original volume limited sample,
but known as planet hosts by the time Paper\,II was published. These stars should, however,
be taken into account for completeness.

\begin{table*}
\caption[]{Stars with planets and derived stellar parameters (HD number from 60\,000 to 160\,000). 
See text for more details.}
\begin{tabular}{lcccrccccc}
\hline
HD     & T$_{\mathrm{eff}}$ & $\log{g}_{spec}$ & $\xi_{\mathrm{t}}$ & \multicolumn{1}{c}{[Fe/H]} & N(\ion{Fe}{i},\ion{Fe}{ii}) & $\sigma$(\ion{Fe}{i},\ion{Fe}{ii}) & Instr.$^a$ & Mass         & $\log{g}_{hipp}$ \\
number & [K]                &  [cm\,s$^{-2}$]  &  [km\,s$^{-1}$]  &        &                              &                                    &            & [M$_{\sun}$] & [cm\,s$^{-2}$] \\
\hline
\object{HD\,65216 }    &  5666$\pm$31  &  4.53$\pm$0.09 &  1.06$\pm$0.05 &$-$0.12$\pm$0.04 & 38,7 & 0.03,0.05 & [2]  & 0.94 &4.53\\
\object{HD\,68988 }    &  5988$\pm$52  &  4.45$\pm$0.15 &  1.25$\pm$0.08 &   0.36$\pm$0.06 & 28,8 & 0.05,0.06 & [4]  & 1.18 &4.41\\
\object{HD\,70642 }    &  5693$\pm$26  &  4.41$\pm$0.09 &  1.01$\pm$0.04 &   0.18$\pm$0.04 & 36,8 & 0.03,0.04 & [1]  & 0.99 &4.43\\
\object{HD\,72659 }    &  5995$\pm$45  &  4.30$\pm$0.07 &  1.42$\pm$0.09 &   0.03$\pm$0.06 & 36,7 & 0.05,0.02 & [2]  & 1.16 &4.22\\
\object{HD\,73256 }    &  5518$\pm$49  &  4.42$\pm$0.12 &  1.22$\pm$0.06 &   0.26$\pm$0.06 & 37,5 & 0.05,0.05 & [2]  & 0.98 &4.51\\
\object{HD\,73526 }    &  5699$\pm$49  &  4.27$\pm$0.12 &  1.26$\pm$0.06 &   0.27$\pm$0.06 & 39,7 & 0.05,0.06 & [2]  & 1.05 &4.15\\
\object{HD\,74156 }    &  6112$\pm$39  &  4.34$\pm$0.10 &  1.38$\pm$0.07 &   0.16$\pm$0.05 & 35,6 & 0.04,0.03 & [2]  & 1.27 &4.16\\
\object{HD\,75289 }    &  6143$\pm$53  &  4.42$\pm$0.13 &  1.53$\pm$0.09 &   0.28$\pm$0.07 & 39,5 & 0.06,0.04 & [1]  & 1.23 &4.35\\
\object{HD\,75732 }    &  5279$\pm$62  &  4.37$\pm$0.18 &  0.98$\pm$0.07 &   0.33$\pm$0.07 & 37,6 & 0.06,0.07 & [3]  & 0.87 &4.44\\
\object{HD\,76700 }    &  5737$\pm$34  &  4.25$\pm$0.14 &  1.18$\pm$0.04 &   0.41$\pm$0.05 & 38,8 & 0.04,0.06 & [2]  & 1.10 &4.26\\
\object{HD\,80606 }    &  5574$\pm$72  &  4.46$\pm$0.20 &  1.14$\pm$0.09 &   0.32$\pm$0.09 & 38,5 & 0.07,0.08 & [3]  & 1.04 &4.55\\
\object{HD\,82943 }    &  6005$\pm$41  &  4.45$\pm$0.13 &  1.08$\pm$0.05 &   0.32$\pm$0.05 & 38,7 & 0.04,0.06 & [1]  & 1.19 &4.41\\
\object{HD\,82943 }    &  6028$\pm$19  &  4.46$\pm$0.02 &  1.18$\pm$0.03 &   0.29$\pm$0.02 & 35,6 & 0.02,0.02 & [5]  & 1.20 &4.43\\
\object{HD\,82943 }    &  6016         &  4.46     	&  1.13          &   \multicolumn{1}{c}{0.30}     	   &      &           & avg. & 1.20 &4.42\\
\object{HD\,83443 }    &  5454$\pm$61  &  4.33$\pm$0.17 &  1.08$\pm$0.08 &   0.35$\pm$0.08 & 38,7 & 0.07,0.08 & [1]  & 0.93 &4.37\\
\object{HD\,89744 }    &  6234$\pm$45  &  3.98$\pm$0.05 &  1.62$\pm$0.08 &   0.22$\pm$0.05 & 26,7 & 0.04,0.02 & [3]  & 1.53 &3.97\\
\object{HD\,92788 }    &  5821$\pm$41  &  4.45$\pm$0.06 &  1.16$\pm$0.05 &   0.32$\pm$0.05 & 37,5 & 0.04,0.02 & [1]  & 1.12 &4.49\\
\object{HD\,95128 }    &  5954$\pm$25  &  4.44$\pm$0.10 &  1.30$\pm$0.04 &   0.06$\pm$0.03 & 30,7 & 0.03,0.04 & [4]  & 1.07 &4.33\\
\object{HD\,106252 }   &  5899$\pm$35  &  4.34$\pm$0.07 &  1.08$\pm$0.06 &$-$0.01$\pm$0.05 & 37,6 & 0.04,0.04 & [1]  & 1.02 &4.39\\
\object{HD\,108147 }   &  6248$\pm$42  &  4.49$\pm$0.16 &  1.35$\pm$0.08 &   0.20$\pm$0.05 & 32,7 & 0.04,0.06 & [1]  & 1.27 &4.41\\
\object{HD\,108874 }   &  5596$\pm$42  &  4.37$\pm$0.12 &  0.89$\pm$0.05 &   0.23$\pm$0.05 & 29,6 & 0.04,0.05 & [3]  & 0.97 &4.27\\
\object{HD\,111232 }   &  5494$\pm$26  &  4.50$\pm$0.10 &  0.84$\pm$0.05 &$-$0.36$\pm$0.04 & 36,6 & 0.03,0.05 & [2]  & 0.75 &4.40\\
\object{HD\,114386 }   &  4804$\pm$61  &  4.36$\pm$0.28 &  0.57$\pm$0.12 &$-$0.08$\pm$0.06 & 35,4 & 0.06,0.14 & [1]  & 0.54 &4.40\\
\object{HD\,114729 }   &  5886$\pm$36  &  4.28$\pm$0.13 &  1.25$\pm$0.09 &$-$0.25$\pm$0.05 & 26,5 & 0.04,0.04 & [3]  & 0.97 &4.13\\
\object{HD\,114762$^b$}&  5884$\pm$34  &  4.22$\pm$0.02 &  1.31$\pm$0.17 &$-$0.70$\pm$0.04 & 34,5 & 0.04,0.02 & [5]  & 0.81 &4.17\\
\object{HD\,114783 }   &  5098$\pm$36  &  4.45$\pm$0.11 &  0.74$\pm$0.05 &   0.09$\pm$0.04 & 27,6 & 0.04,0.05 & [4]  & 0.77 &4.52\\
\object{HD\,117176 }   &  5560$\pm$34  &  4.07$\pm$0.05 &  1.18$\pm$0.05 &$-$0.06$\pm$0.05 & 33,6 & 0.04,0.02 & [4]  & 0.93 &3.87\\
\object{HD\,120136 }   &  6339$\pm$73  &  4.19$\pm$0.10 &  1.70$\pm$0.16 &   0.23$\pm$0.07 & 24,4 & 0.05,0.04 & [5]  & 1.33 &4.25\\
\object{HD\,121504 }   &  6075$\pm$40  &  4.64$\pm$0.12 &  1.31$\pm$0.07 &   0.16$\pm$0.05 & 39,7 & 0.04,0.05 & [1]  & 1.17 &4.41\\
\object{HD\,128311 }   &  4835$\pm$72  &  4.44$\pm$0.21 &  0.89$\pm$0.11 &   0.03$\pm$0.07 & 26,5 & 0.07,0.09 & [3]  & 0.61 &4.43\\
\object{HD\,130322 }   &  5392$\pm$36  &  4.48$\pm$0.06 &  0.85$\pm$0.05 &   0.03$\pm$0.04 & 32,6 & 0.04,0.03 & [4]  & 0.96 &4.61\\
\object{HD\,134987 }   &  5776$\pm$29  &  4.36$\pm$0.07 &  1.09$\pm$0.04 &   0.30$\pm$0.04 & 31,7 & 0.03,0.03 & [4]  & 1.08 &4.32\\
\object{HD\,136118$^b$}&  6222$\pm$39  &  4.27$\pm$0.15 &  1.79$\pm$0.12 &$-$0.04$\pm$0.05 & 27,7 & 0.03,0.06 & [4]  & 1.29 &4.12\\
\object{HD\,137759 }   &  4775$\pm$113 &  3.09$\pm$0.40 &  1.78$\pm$0.11 &   0.13$\pm$0.14 & 29,7 & 0.12,0.18 & [4]  &  --  & -- \\
\object{HD\,141937 }   &  5909$\pm$39  &  4.51$\pm$0.08 &  1.13$\pm$0.06 &   0.10$\pm$0.05 & 38,7 & 0.04,0.03 & [3]  & 1.08 &4.45\\
\object{HD\,142415 }   &  6045$\pm$44  &  4.53$\pm$0.08 &  1.12$\pm$0.07 &   0.21$\pm$0.05 & 38,7 & 0.05,0.04 & [2]  & 1.26 &4.57\\
\object{HD\,143761 }   &  5853$\pm$25  &  4.41$\pm$0.15 &  1.35$\pm$0.07 &$-$0.21$\pm$0.04 & 31,6 & 0.03,0.06 & [4]  & 0.95 &4.20\\
\object{HD\,145675 }   &  5311$\pm$87  &  4.42$\pm$0.18 &  0.92$\pm$0.10 &   0.43$\pm$0.08 & 29,5 & 0.06,0.05 & [4]  & 0.90 &4.41\\
\object{HD\,147513 }   &  5883$\pm$25  &  4.51$\pm$0.05 &  1.18$\pm$0.04 &   0.06$\pm$0.04 & 36,7 & 0.03,0.03 & [1]  & 1.11 &4.53\\
\object{HD\,150706 }   &  5961$\pm$27  &  4.50$\pm$0.10 &  1.11$\pm$0.06 &$-$0.01$\pm$0.04 & 27,5 & 0.03,0.05 & [3]  & 1.17 &4.59\\
\hline
\end{tabular}
\\ $^a$ The instruments used to obtain the spectra were: [1] 1.2-m Swiss Telescope/CORALIE; [2] 1.5-m and 2.2-m ESO/FEROS; [3] WHT/UES; [4] TNG/SARG; [5] VLT-UT2/UVES; [6] 1.93-m OHP/ELODIE; [7] Keck/HIRES
\newline
$^b$ The companions to these stars have minimum masses above 10\,M$_{\mathrm{Jup}}$, and are probably Brown-Dwarfs.  
\label{tabplan2}
\end{table*}

\begin{table*}
\caption[]{Stars with planets and derived stellar parameters (HD number from 160\,000 on). 
See text for more details.}
\begin{tabular}{lcccrccccc}
\hline
HD     & T$_{\mathrm{eff}}$ & $\log{g}_{spec}$ & $\xi_{\mathrm{t}}$ & \multicolumn{1}{c}{[Fe/H]} & N(\ion{Fe}{i},\ion{Fe}{ii}) & $\sigma$(\ion{Fe}{i},\ion{Fe}{ii}) & Instr.$^a$ & Mass         & $\log{g}_{hipp}$ \\
number & [K]                &  [cm\,s$^{-2}$]  &  [km\,s$^{-1}$]  &        &                              &                                    &            & [M$_{\sun}$] & [cm\,s$^{-2}$] \\
\hline
\object{HD\,160691 }   &  5798$\pm$33  &  4.31$\pm$0.08 &  1.19$\pm$0.04 &   0.32$\pm$0.04 & 36,7 & 0.04,0.03 & [1]  & 1.10 &4.25\\
\object{HD\,162020$^b$}&  4835$\pm$72  &  4.39$\pm$0.25 &  0.86$\pm$0.12 &$-$0.09$\pm$0.07 & 36,4 & 0.07,0.1  & [1]  & 0.66 &4.56\\
\object{HD\,162020$^b$}&  4882$\pm$91  &  4.44$\pm$0.35 &  0.87$\pm$0.16 &   0.01$\pm$0.08 & 35,4 & 0.08,0.18 & [2]  & 0.80 &4.67\\
\object{HD\,162020$^b$}&  4858         &  4.42     	&  0.86          &\multicolumn{1}{c}{$-$0.04}     	   &      &           & avg. & 0.73 &4.62\\
\object{HD\,168443 }   &  5617$\pm$35  &  4.22$\pm$0.05 &  1.21$\pm$0.05 &   0.06$\pm$0.05 & 31,7 & 0.04,0.02 & [4]  & 0.96 &4.05\\
\object{HD\,168746 }   &  5601$\pm$33  &  4.41$\pm$0.12 &  0.99$\pm$0.05 &$-$0.08$\pm$0.05 & 38,7 & 0.04,0.05 & [1]  & 0.88 &4.31\\
\object{HD\,169830 }   &  6299$\pm$41  &  4.10$\pm$0.02 &  1.42$\pm$0.09 &   0.21$\pm$0.05 & 38,4 & 0.04,0.01 & [1]  & 1.43 &4.09\\
\object{HD\,177830 }   &  4804$\pm$77  &  3.57$\pm$0.17 &  1.14$\pm$0.09 &   0.33$\pm$0.09 & 31,4 & 0.08,0.04 & [4]  &  --  &--  \\
\object{HD\,178911B$^c$}& 5588$\pm$115 &  4.46$\pm$0.20 &  0.82$\pm$0.14 &   0.24$\pm$0.10 & 16,2 & 0.06,0.02 & [7]  & 0.97 &3.73\\
\object{HD\,178911B}   &  5600$\pm$42  &  4.44$\pm$0.08 &  0.95$\pm$0.05 &   0.27$\pm$0.05 & 30,6 & 0.04,0.04 & [4]  & 0.98 &3.74\\
\object{HD\,179949 }   &  6260$\pm$43  &  4.43$\pm$0.05 &  1.41$\pm$0.09 &   0.22$\pm$0.05 & 34,5 & 0.04,0.02 & [1]  & 1.28 &4.43\\
\object{HD\,186427 }   &  5772$\pm$25  &  4.40$\pm$0.07 &  1.07$\pm$0.04 &   0.08$\pm$0.04 & 33,7 & 0.03,0.02 & [4]  & 0.99 &4.35\\
\object{HD\,187123 }   &  5845$\pm$22  &  4.42$\pm$0.07 &  1.10$\pm$0.03 &   0.13$\pm$0.03 & 30,6 & 0.02,0.03 & [4]  & 1.04 &4.33\\
\object{HD\,190228 }   &  5312$\pm$30  &  3.87$\pm$0.05 &  1.11$\pm$0.04 &$-$0.25$\pm$0.05 & 35,7 & 0.04,0.02 & [4]  &  --  &--  \\
\object{HD\,190228 }   &  5342$\pm$39  &  3.93$\pm$0.09 &  1.11$\pm$0.05 &$-$0.27$\pm$0.06 & 37,6 & 0.05,0.04 & [3]  &  --  &--  \\
\object{HD\,190228 }   &  5327         &  3.90     	&  1.11     	 &\multicolumn{1}{c}{$-$0.26}     	   &      &           & avg. &  --  &--  \\ 
\object{HD\,190360A}   &  5584$\pm$36  &  4.37$\pm$0.06 &  1.07$\pm$0.05 &   0.24$\pm$0.05 & 29,5 & 0.04,0.02 & [3]  & 0.96 &4.32\\
\object{HD\,192263 }   &  4947$\pm$58  &  4.51$\pm$0.20 &  0.86$\pm$0.09 &$-$0.02$\pm$0.06 & 35,6 & 0.06,0.10 & [2]  & 0.69 &4.51\\
\object{HD\,195019A}   &  5859$\pm$31  &  4.32$\pm$0.07 &  1.27$\pm$0.05 &   0.09$\pm$0.04 & 39,7 & 0.04,0.03 & [1]  & 1.06 &4.18\\
\object{HD\,195019A}   &  5836$\pm$39  &  4.31$\pm$0.07 &  1.27$\pm$0.06 &   0.06$\pm$0.05 & 35,7 & 0.04,0.03 & [4]  & 1.05 &4.16\\
\object{HD\,195019A}   &  5842         &  4.32     	&  1.27          &   \multicolumn{1}{c}{0.08}     	   &      &           & avg. & 1.06 &4.17\\
\object{HD\,196050 }   &  5918$\pm$44  &  4.35$\pm$0.13 &  1.39$\pm$0.06 &   0.22$\pm$0.05 & 36,7 & 0.04,0.05 & [1]  & 1.15 &4.29\\
\object{HD\,202206$^b$}&  5752$\pm$53  &  4.50$\pm$0.09 &  1.01$\pm$0.06 &   0.35$\pm$0.06 & 39,6 & 0.05,0.04 & [1]  & 1.06 &4.43\\
\object{HD\,209458 }   &  6117$\pm$26  &  4.48$\pm$0.08 &  1.40$\pm$0.06 &   0.02$\pm$0.03 & 34,7 & 0.02,0.03 & [5]  & 1.15 &4.41\\
\object{HD\,210277 }   &  5546$\pm$28  &  4.29$\pm$0.09 &  1.06$\pm$0.03 &   0.21$\pm$0.04 & 36,6 & 0.04,0.04 & [2]  & 0.94 &4.36\\
\object{HD\,210277 }   &  5519$\pm$26  &  4.29$\pm$0.18 &  1.01$\pm$0.03 &   0.16$\pm$0.04 & 34,7 & 0.03,0.08 & [4]  & 0.91 &4.34\\
\object{HD\,210277}    &  5532         &  4.29     	&  1.04          &   \multicolumn{1}{c}{0.19}     	   &      &           & avg. & 0.92 &4.35\\
\object{HD\,213240 }   &  5984$\pm$33  &  4.25$\pm$0.10 &  1.25$\pm$0.05 &   0.17$\pm$0.05 & 38,7 & 0.04,0.04 & [1]  & 1.22 &4.18\\
\object{HD\,216435 }   &  5938$\pm$42  &  4.12$\pm$0.05 &  1.28$\pm$0.06 &   0.24$\pm$0.05 & 33,6 & 0.04,0.03 & [1]  & 1.34 &4.07\\
\object{HD\,216437 }   &  5887$\pm$32  &  4.30$\pm$0.07 &  1.31$\pm$0.04 &   0.25$\pm$0.04 & 37,7 & 0.03,0.03 & [1]  & 1.20 &4.21\\
\object{HD\,216770 }   &  5423$\pm$41  &  4.40$\pm$0.13 &  1.01$\pm$0.05 &   0.26$\pm$0.04 & 30,7 & 0.04,0.07 & [4]  & 0.91 &4.42\\
\object{HD\,217014 }   &  5804$\pm$36  &  4.42$\pm$0.07 &  1.20$\pm$0.05 &   0.20$\pm$0.05 & 35,6 & 0.04,0.02 & [2]  & 1.05 &4.36\\
\object{HD\,217107 }   &  5630$\pm$32  &  4.28$\pm$0.12 &  1.02$\pm$0.04 &   0.37$\pm$0.05 & 38,7 & 0.04,0.05 & [1]  & 1.01 &4.34\\
\object{HD\,217107 }   &  5663$\pm$36  &  4.34$\pm$0.08 &  1.11$\pm$0.04 &   0.37$\pm$0.05 & 37,7 & 0.04,0.03 & [2]  & 1.02 &4.36\\
\object{HD\,217107 }   &  5646         &  4.31     	&  1.06          &   \multicolumn{1}{c}{0.37}     	   &      &           & avg. & 1.02 &4.35\\
\object{HD\,219542B}   &  5732$\pm$31  &  4.40$\pm$0.05 &  0.99$\pm$0.04 &   0.17$\pm$0.04 & 32,7 & 0.03,0.03 & [4]  & 1.04 &4.08\\
\object{HD\,222404 }   &  4916$\pm$70  &  3.36$\pm$0.21 &  1.27$\pm$0.06 &   0.16$\pm$0.08 & 26,7 & 0.07,0.08 & [4]  &  --  &--  \\
\object{HD\,222582 }   &  5843$\pm$38  &  4.45$\pm$0.07 &  1.03$\pm$0.06 &   0.05$\pm$0.05 & 36,7 & 0.04,0.03 & [3]  & 1.02 &4.38\\
\hline
\end{tabular}
\\ $^a$ The instruments used to obtain the spectra were: [1] 1.2-m Swiss Telescope/CORALIE; [2] 1.5-m and 2.2-m ESO/FEROS; [3] WHT/UES; [4] TNG/SARG; [5] VLT-UT2/UVES; [6] 1.93-m OHP/ELODIE; [7] Keck/HIRES
\newline
$^b$ The companions to these stars have minimum masses above 10\,M$_{\mathrm{Jup}}$, , and are probably Brown-Dwarfs..  
\newline
$^c$ These parameters, derived from a Keck/HIRES spectrum, were computed with a reduced 
number of iron lines. In the rest of the paper, only the parameters derived 
from the SARG/TNG spectrum were considered.
\label{tabplan3}
\end{table*}

\section{Spectroscopic analysis and stellar parameters}

For the past three years we have been deriving stellar parameters for planet-host stars 
and for a comparison sample of stars with no detected planetary companions (Papers I, II and III).
However, the stellar parameters
presented in our previous studies were not completely satisfactory. In particular, 
the derived surface gravities were systematically higher than the ones obtained
by other authors \citep[see e.g.][]{Gon01} by $\sim$0.15\,dex. While this fact 
was clearly not producing an important shift in the final metallicities \citep[see e.g.][]{San03,Law03}, 
this lead some authors to suggest that the metallicity excess observed was not 
real (G. Wuchterl, private communication).

To solve this problem we have carried out a new spectroscopic analysis
of all the program stars. The stellar parameters were derived using the same
technique as in the previous papers, based on about 39 \ion{Fe}{i} and 12 \ion{Fe}{ii} 
lines (see Table\,\ref{tab1}), and the spectroscopic analysis was done in LTE using the 
2002 version of the code MOOG \citep[][]{Sne73}\footnote{The code MOOG2002 can be downloaded at http://verdi.as.utexas.edu/moog.html}. However, 2 main changes have been
done. Firstly, we have adopted new $\log{gf}$ values for the iron lines. These were
computed from an inverted solar analysis using solar EW measured from
the Kurucz Solar Atlas \citep[][]{Kur84}, and a Kurucz grid model for the Sun \citep[][]{Kur93} having (T$_{\mathrm{eff}}$,$\log{g}$,$\xi_{\mathrm{t}}$,$log{\epsilon}_{Fe}$)=(5777K,4.44\,dex,1.00\,km\,s$^{-1}$,7.47\,dex).
This differs from our previous analysis where we always used $\log{gf}$ values taken
from \citet[][]{Gon01} (and references therein).
Secondly, we have now used a van der Walls damping based on the Unsold approximation, but
multiplied by a factor as suggested by the Blackwell group (option 2 in the 
damping parameter inside MOOG). 

We also note that our previous analysis was done using an older version of
MOOG. A comparison showed that for some cases there were slight differences
in the derived stellar metallicities, but never exceeding 0.01\,dex.

As a test, we computed the Solar parameters and iron abundances
based on iron EW measured using a Solar spectrum taken with the HARPS spectrograph (courtesy of 
the HARPS team, Mayor et al.). The resulting parameters were T$_{\mathrm{eff}}$=5779$\pm$23, $\log{g}$=4.48$\pm$0.07, $\xi_{\mathrm{t}}$=1.04$\pm$0.04, and [Fe/H]=0.00$\pm$0.03, very close (and within the errors) 
to the ``expected'' solution (there are almost no differences in average 
between the solar EW derived from the Kurucz Atlas compared to the ones derived from the HARPS spectrum).

The atmospheric parameters for our program stars were obtained from the \ion{Fe}{i} and \ion{Fe}{ii} lines
by iterating until the correlation coefficients between $\log{\epsilon}$(\ion{Fe}{i}) 
and $\chi_l$, and between $\log{\epsilon}$(\ion{Fe}{i}) and  $\log{({W}_\lambda/\lambda)}$ 
were zero, and the mean abundance given by \ion{Fe}{i} and \ion{Fe}{ii} lines were the same. 
To simplify this analysis, we built a Fortran code that uses a Downhill Simplex Method 
\citep[][]{Numerical} to find the best solution in the (stellar) parameter space 
(which happens in most of the cases after a few minutes). The results are thus obtained in a 
fast and automatic way, once the EW are measured.		   

The final stellar parameters and masses are presented in Tables\,\ref{tabplan1} through \ref{tabcomp},
for planet-host stars and for our comparison sample objects\footnote{These tables are also available
in electronic form at CDS}. The errors were derived as described in Paper\,I, and are of the order
of 50\,K in T$_{\mathrm{eff}}$, 0.12\,dex in $\log{g}$, 0.08\,km\,s$^{-1}$ in the microturbulence, 
and 0.05\,dex in the metallicity. Stellar masses were 
computed by interpolating the theoretical isochrones of \citet{Sch92}, 
\citet[][]{Sch93} and \citet{Schae92}, using $M_{V}$ computed using Hipparcos 
parallaxes \citep[][]{ESA97}, a bolometric correction from \citet[][]{Flo96}, and the
T$_{\mathrm{eff}}$ obtained from the spectroscopy. 
We adopt a typical relative error of 0.05\,M$_{\sun}$ for the masses. In some cases,
no mass estimates are presented, since these involved large extrapolations of the
isochrones. {A comparison with other works shows that (on average) there are almost no 
differences to the masses derived in the study of \citet[][]{Law03}, although these
authors used a different set of theoretical isochrones;
a small difference of 0.03\,M$_{\sun}$ is found with respect to the
analysis of \citet[][]{ALP99}.}

\begin{table*}
\caption[]{List of 41 stars from our comparison sample and derived stellar parameters. See text for more details.}
\begin{tabular}{lcccrccccc}
\hline
HD     & T$_{\mathrm{eff}}$ & $\log{g}_{spec}$ & $\xi_{\mathrm{t}}$ & \multicolumn{1}{c}{[Fe/H]} & N(\ion{Fe}{i},\ion{Fe}{ii}) & $\sigma$(\ion{Fe}{i},\ion{Fe}{ii}) & Instr.$^a$ & Mass         & $\log{g}_{hipp}$ \\
number & [K]                &  [cm\,s$^{-2}$]  &  [km\,s$^{-1}$]  &        &                              &                                    &            & [M$_{\sun}$] & [cm\,s$^{-2}$] \\
\hline
\object{HD\,1581  } & 5956$\pm$44 & 4.39$\pm$0.13 & 1.07$\pm$0.09 & $-$0.14$\pm$0.05 & 33,7 & 0.04,0.05 & [2] & 1.00 &4.41\\
\object{HD\,4391  } & 5878$\pm$53 & 4.74$\pm$0.15 & 1.13$\pm$0.10 & $-$0.03$\pm$0.06 & 35,5 & 0.05,0.05 & [1] & 1.11 &4.57\\
\object{HD\,5133  } & 4911$\pm$54 & 4.49$\pm$0.18 & 0.71$\pm$0.11 & $-$0.17$\pm$0.06 & 38,6 & 0.06,0.09 & [1] & 0.63 &4.49\\
\object{HD\,7570  } & 6140$\pm$41 & 4.39$\pm$0.16 & 1.50$\pm$0.08 &    0.18$\pm$0.05 & 35,6 & 0.04,0.05 & [1] & 1.20 &4.36\\
\object{HD\,10360 } & 4970$\pm$40 & 4.49$\pm$0.10 & 0.76$\pm$0.07 & $-$0.26$\pm$0.04 & 37,5 & 0.05,0.05 & [1] & 0.62 &4.44\\
\object{HD\,10700 } & 5344$\pm$29 & 4.57$\pm$0.09 & 0.91$\pm$0.06 & $-$0.52$\pm$0.04 & 38,6 & 0.03,0.04 & [1] & 0.65 &4.43\\
\object{HD\,14412 } & 5368$\pm$24 & 4.55$\pm$0.05 & 0.88$\pm$0.05 & $-$0.47$\pm$0.03 & 35,6 & 0.03,0.02 & [1] & 0.73 &4.54\\
\object{HD\,17925 } & 5180$\pm$56 & 4.44$\pm$0.13 & 1.33$\pm$0.08 &    0.06$\pm$0.07 & 35,6 & 0.06,0.06 & [1] & 0.84 &4.58\\
\object{HD\,20010 } & 6275$\pm$57 & 4.40$\pm$0.37 & 2.41$\pm$0.41 & $-$0.19$\pm$0.06 & 33,7 & 0.05,0.14 & [1] & 1.33 &4.03\\
\object{HD\,20766 } & 5733$\pm$31 & 4.55$\pm$0.10 & 1.09$\pm$0.06 & $-$0.21$\pm$0.04 & 37,7 & 0.03,0.04 & [1] & 0.93 &4.51\\
\object{HD\,20794 } & 5444$\pm$31 & 4.47$\pm$0.07 & 0.98$\pm$0.06 & $-$0.38$\pm$0.04 & 39,6 & 0.04,0.03 & [1] & 0.72 &4.38\\
\object{HD\,20807 } & 5843$\pm$26 & 4.47$\pm$0.10 & 1.17$\pm$0.06 & $-$0.23$\pm$0.04 & 37,7 & 0.03,0.04 & [1] & 0.94 &4.45\\
\object{HD\,23249 } & 5074$\pm$60 & 3.77$\pm$0.16 & 1.08$\pm$0.06 &    0.13$\pm$0.08 & 38,5 & 0.07,0.07 & [1] & --   &--  \\
\object{HD\,23356 } & 4975$\pm$55 & 4.48$\pm$0.16 & 0.77$\pm$0.09 & $-$0.11$\pm$0.06 & 38,6 & 0.06,0.09 & [1] & 0.71 &4.57\\
\object{HD\,23484 } & 5176$\pm$45 & 4.41$\pm$0.17 & 1.03$\pm$0.06 &    0.06$\pm$0.05 & 38,6 & 0.05,0.08 & [1] & 0.82 &4.55\\
\object{HD\,26965A} & 5126$\pm$34 & 4.51$\pm$0.08 & 0.60$\pm$0.07 & $-$0.31$\pm$0.04 & 38,5 & 0.04,0.04 & [1] & 0.65 &4.42\\
\object{HD\,30495 } & 5868$\pm$30 & 4.55$\pm$0.10 & 1.24$\pm$0.05 &    0.02$\pm$0.04 & 37,7 & 0.03,0.04 & [1] & 1.10 &4.54\\
\object{HD\,36435 } & 5479$\pm$37 & 4.61$\pm$0.07 & 1.12$\pm$0.05 & $-$0.00$\pm$0.05 & 38,6 & 0.04,0.04 & [1] & 0.98 &4.60\\
\object{HD\,38858 } & 5752$\pm$32 & 4.53$\pm$0.07 & 1.26$\pm$0.07 & $-$0.23$\pm$0.05 & 37,7 & 0.03,0.02 & [1] & 0.91 &4.47\\
\object{HD\,40307 } & 4805$\pm$52 & 4.37$\pm$0.37 & 0.49$\pm$0.12 & $-$0.30$\pm$0.05 & 37,5 & 0.06,0.20 & [1] & --   &--  \\
\object{HD\,43162 } & 5633$\pm$35 & 4.48$\pm$0.07 & 1.24$\pm$0.05 & $-$0.01$\pm$0.04 & 34,6 & 0.04,0.03 & [1] & 1.00 &4.57\\
\object{HD\,43834 } & 5594$\pm$36 & 4.41$\pm$0.09 & 1.05$\pm$0.04 &    0.10$\pm$0.05 & 38,5 & 0.04,0.04 & [1] & 0.93 &4.44\\
\object{HD\,50281A} & 4658$\pm$56 & 4.32$\pm$0.24 & 0.64$\pm$0.15 & $-$0.04$\pm$0.07 & 34,4 & 0.06,0.12 & [1] & --   &--  \\
\object{HD\,53705 } & 5825$\pm$20 & 4.37$\pm$0.10 & 1.20$\pm$0.04 & $-$0.19$\pm$0.03 & 36,7 & 0.02,0.03 & [1] & 0.93 &4.31\\
\object{HD\,53706 } & 5260$\pm$31 & 4.35$\pm$0.11 & 0.74$\pm$0.05 & $-$0.26$\pm$0.04 & 35,6 & 0.04,0.05 & [1] & 0.78 &4.57\\
\object{HD\,65907A} & 5979$\pm$31 & 4.59$\pm$0.12 & 1.36$\pm$0.10 & $-$0.29$\pm$0.04 & 38,7 & 0.03,0.05 & [1] & 0.96 &4.39\\
\object{HD\,69830 } & 5410$\pm$26 & 4.38$\pm$0.07 & 0.89$\pm$0.03 & $-$0.03$\pm$0.04 & 38,7 & 0.03,0.04 & [1] & 0.84 &4.48\\
\object{HD\,72673 } & 5242$\pm$28 & 4.50$\pm$0.09 & 0.69$\pm$0.05 & $-$0.37$\pm$0.04 & 38,6 & 0.03,0.05 & [1] & 0.71 &4.53\\
\object{HD\,74576 } & 5000$\pm$55 & 4.55$\pm$0.13 & 1.07$\pm$0.08 & $-$0.03$\pm$0.06 & 37,5 & 0.06,0.06 & [1] & 0.78 &4.62\\
\object{HD\,76151 } & 5803$\pm$29 & 4.50$\pm$0.08 & 1.02$\pm$0.04 &    0.14$\pm$0.04 & 39,7 & 0.03,0.05 & [1] & 1.07 &4.50\\
\object{HD\,84117 } & 6167$\pm$37 & 4.35$\pm$0.10 & 1.42$\pm$0.09 & $-$0.03$\pm$0.05 & 35,5 & 0.04,0.04 & [1] & 1.15 &4.34\\
\object{HD\,189567} & 5765$\pm$24 & 4.52$\pm$0.05 & 1.22$\pm$0.05 & $-$0.23$\pm$0.04 & 37,5 & 0.03,0.02 & [1] & 0.89 &4.39\\
\object{HD\,191408A}& 5005$\pm$45 & 4.38$\pm$0.25 & 0.67$\pm$0.09 & $-$0.55$\pm$0.06 & 38,4 & 0.05,0.12 & [1] & --   &--  \\
\object{HD\,192310} & 5069$\pm$49 & 4.38$\pm$0.19 & 0.79$\pm$0.07 & $-$0.01$\pm$0.05 & 36,6 & 0.05,0.09 & [1] & 0.72 &4.47\\
\object{HD\,196761} & 5435$\pm$39 & 4.48$\pm$0.08 & 0.91$\pm$0.07 & $-$0.29$\pm$0.05 & 38,5 & 0.04,0.04 & [1] & 0.78 &4.49\\
\object{HD\,207129} & 5910$\pm$24 & 4.42$\pm$0.05 & 1.14$\pm$0.04 &    0.00$\pm$0.04 & 37,6 & 0.03,0.02 & [1] & 1.04 &4.42\\
\object{HD\,209100} & 4629$\pm$77 & 4.36$\pm$0.19 & 0.42$\pm$0.25 & $-$0.06$\pm$0.08 & 36,3 & 0.07,0.06 & [1] & --   &--  \\
\object{HD\,211415} & 5890$\pm$30 & 4.51$\pm$0.07 & 1.12$\pm$0.07 & $-$0.17$\pm$0.04 & 35,7 & 0.03,0.02 & [1] & 0.97 &4.42\\
\object{HD\,216803} & 4555$\pm$87 & 4.53$\pm$0.26 & 0.66$\pm$0.28 & $-$0.01$\pm$0.09 & 30,3 & 0.08,0.10 & [1] & --   &--  \\
\object{HD\,222237} & 4747$\pm$58 & 4.48$\pm$0.22 & 0.40$\pm$0.20 & $-$0.31$\pm$0.06 & 37,4 & 0.07,0.11 & [1] & --   &--  \\
\object{HD\,222335} & 5260$\pm$41 & 4.45$\pm$0.11 & 0.92$\pm$0.06 & $-$0.16$\pm$0.05 & 35,6 & 0.04,0.05 & [2] & 0.77 &4.52\\
\hline
\end{tabular}
\\ $^a$ The instruments used to obtain the spectra were: [1] CORALIE; [2] FEROS
\label{tabcomp}
\end{table*}

\begin{figure}[t]
\psfig{width=\hsize,file=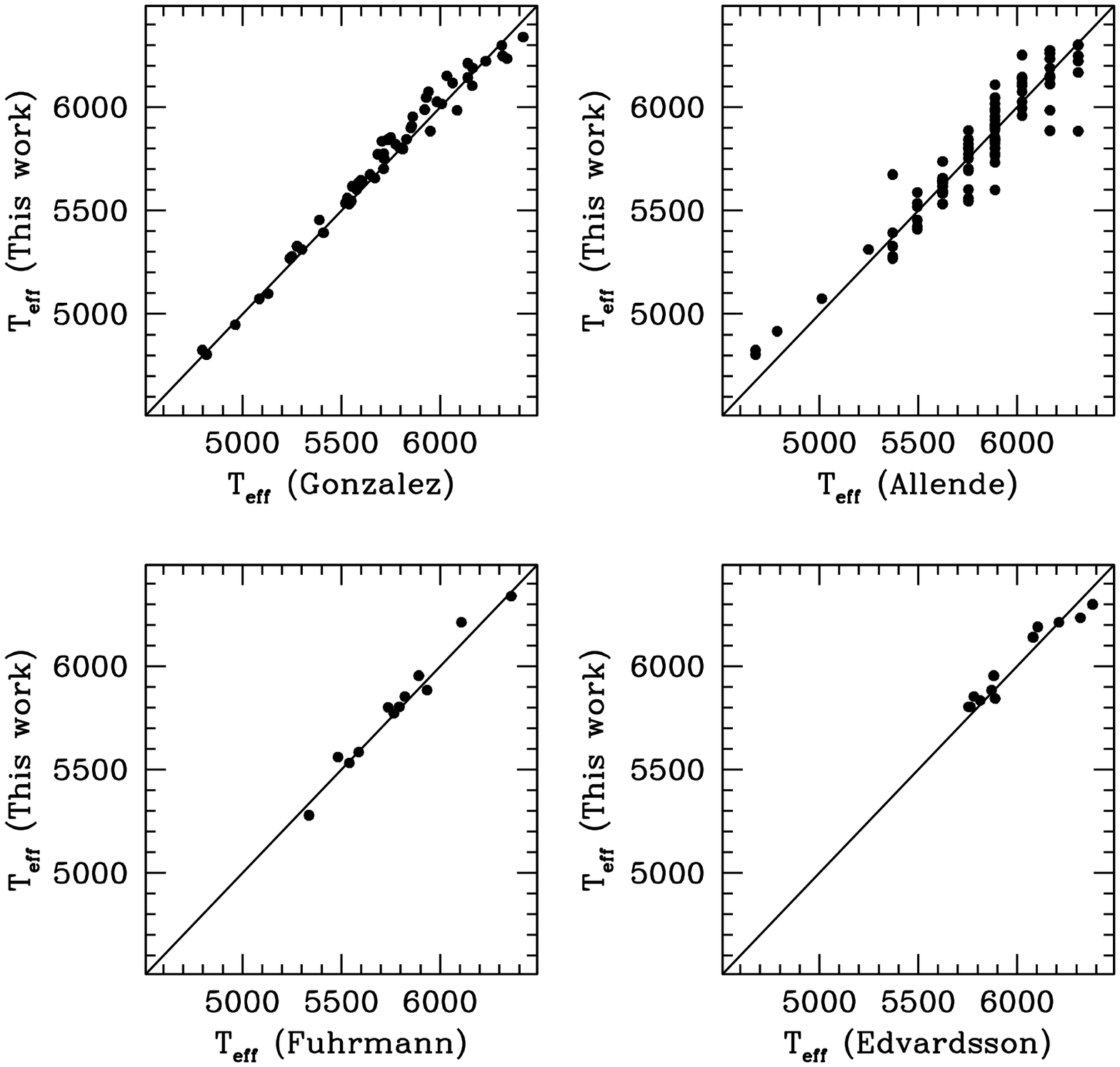}
\caption[]{Comparison of the T$_{\mathrm{eff}}$ values derived in this
work with the ones obtained by other authors for the same stars. The solid line represents a
1:1 relation. See text for more details.}
\label{figteff}
\end{figure}

For comparison, we have also computed 
the surface gravities based on Hipparcos parallaxes (trigonometric gravities). Using the
well known relations $g\,=\,\frac{GM}{R^{2}}$, and $L\,=\,\frac{4}{3}\pi\,R^{2}T^{4}_{\mathrm{eff}}$, we can obtain:
\begin{eqnarray}
\label{eqnlogg}
\log{\frac{g}{g_{\sun}}} & = & \log{\frac{M}{M_{\sun}}}+4\log{\frac{T_{\mathrm{eff}}}{T_{\mathrm{eff}{\sun}}}}+2\log{\pi} \\
                 &   & +0.4(V_{0}+BC)+0.11\nonumber 
\end{eqnarray}
where BC is the bolometric correction, V$_{0}$ the visual magnitude, and $\pi$ the parallax.
Here we used a solar absolute magnitude M$_{v}$=4.81 \citep[][]{Bes98} and, for
consistency, we took the bolometric
correction derived for a solar temperature star ($-$0.08) using the calibration of 
\citet[][]{Flo96}\footnote{We can find some differences in the literature regarding these 
values \citep[see e.g.][]{Bes98,Ber92}, which can introduce systematic errors in the resulting trigonometric 
parallaxes. In particular, there seems to be a large
discrepancy regarding the solar BC derived using Kurucz models \citep[see ][]{Bes98}}. 
This method was already successfully used by other authors, namely \citet[][]{All99} and \citet[][]{Nis97}, 
in obtaining surface gravities for stars with precise parallax estimates. 
Given the proximity of our targets (typical values of $\sigma(\pi)/\pi$ are 
lower than 0.05, and always lower than 0.10 except for \object{HD\,80606}), the derived 
trigonometric surface gravities are reasonably free from the Lutz-Kelker 
effect \citep[][]{Lut73,Smi03}. In the next section we will present the results of a comparison
between our spectroscopic and trigonometric gravities.

Finally, for a few stars we have stellar parameters and metallicities derived using
different sets of spectra. A simple inspection of Tables\,\ref{tabplan1}, \ref{tabplan2}, and \ref{tabplan3} 
shows that the parameters derived from these different spectra are perfectly compatible with each other, 
within the errors.

\subsection{Comparison with other works}

To verify the quality of our results we have made a comparison with a number of different 
studies.
In particular, we have compared the presented stellar parameters with the ones derived in
our previous works (Papers\,II and III). This comparison reveals one main difference: the
derived values for the surface gravity are now lower by about $\sim$0.1 dex (on average). 
However, for both the effective temperatures and metallicities, 
the differences are very small, not exceeding $\sim$10\,k and 0.01\,dex, respectively.
In other words, the new parameters do not differ considerably in the
main goal of our studies: the derivation of precise [Fe/H]. The changes we have made have not produced 
much of a difference in the obtained metallicities. This 
conclusion was expected, as it is well known that for solar-type dwarfs the abundances derived from 
the \ion{Fe}{i} lines are mostly sensitive to the effective temperature (that did not vary much from our 
previous analysis to the current one) and are almost not dependent on surface gravity 
variations (see e.g. Paper I).

To verify this case we have performed a test where we used the solar equivalent widths (used to
derive the $\log{gf}$ values) to obtain the effective temperature, microturbulence parameter,
and metallicity for the Sun based only on the \ion{Fe}{i} lines, and forcing 
the $\log{g}$ to a value of 4.54\,dex, i.e., 0.1\,dex above solar. The results were 
(T$_{\mathrm{eff}}$, $\xi_{\mathrm{t}}$, [Fe/H])=(5755\,K, 0.94\,km\.s$^{-1}$, $-$0.01\,dex), 
not very different to the ``expected'' solar values. Similar or lower 
differences were obtained on a test done for the hotter dwarfs \object{HD\,82943} 
and \object{HD\,84177}.

A clear conclusion of this analysis is that the method we used to derive stellar 
metallicities is not very dependent on errors in $\log{g}$. 

\subsubsection{Effective Temperatures}

We have further compared our stellar parameters with the ones derived by other 
authors for the stars in common.

For the T$_{\mathrm{eff}}$ we have found that our values are only $+$18\,k in excess of 
those derived in the works of \citet[][]{Fur97}, \citet[][]{Fur98}, and \citet[][]{Furetal98}, 
who used a H$_\alpha$ 
and H$_\beta$ line-fitting procedure to derive the effective temperatures (we have 12 stars 
in common) -- see Fig.\,\ref{figteff}. Similarly, a small average difference of $+$25\,k is found to the 
studies of \citet[][]{Gon01}, \citet[][]{Law03}, and references therein (57 stars; using a similar 
technique to ours), of $+$16\,K to \citet[][]{Edv93} (12 stars; T$_{\mathrm{eff}}$ 
derived from photometry), and of $-$2\,K to \citet[][]{ALP99} (90 stars; these authors 
used an evolutionary model-fitting procedure to derive the stellar parameters). 
{An insignificant average difference of 3\,K is also found
when comparing our results with the values obtained by \citet[][]{Rib03}, 
based on IR photometry.}

\begin{figure}[t]
\psfig{width=\hsize,file=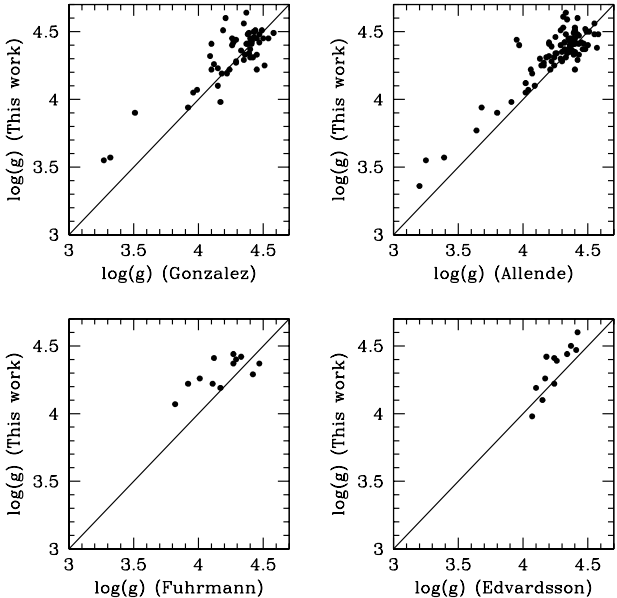}
\caption[]{Comparison of the spectroscopic $\log{g}$ values derived in this
work with the ones obtained by other authors for the same stars. The solid line represents a
1:1 relation. See text for more details.}
\label{figlogg}
\end{figure}

\subsubsection{Surface gravities}

For surface gravities, we have also found small differences to the other 
studies, (when compared with the individual errors or the order of 0.12\,dex) -- see Fig.\,\ref{figlogg}.
In particular, our $\log{g}$s are only $\sim$0.05\,dex (on average) above the ones derived by
\citet[][]{Gon01}, \citet[][]{Law03}, and references therein, and 0.08\,dex above the 
results of \citet[][]{ALP99}, and \citet[][]{Edv93} (i.e., differences of the order of 1-2\%). 
This difference is even smaller (below 0.04\,dex, when compared with the results of the 
Gonzalez group) if we do not consider the most evolved stars. 
A slightly higher difference of about $+$0.10\,dex is also
found to the works of \citet[][]{Fur97}, \citet[][]{Fur98} and \citet[][]{Furetal98}; these authors had already
found that their spectroscopic gravities were lower than trigonometric-based parallaxes
by about 0.03\,dex.
 
\begin{figure}[t]
\psfig{width=\hsize,file=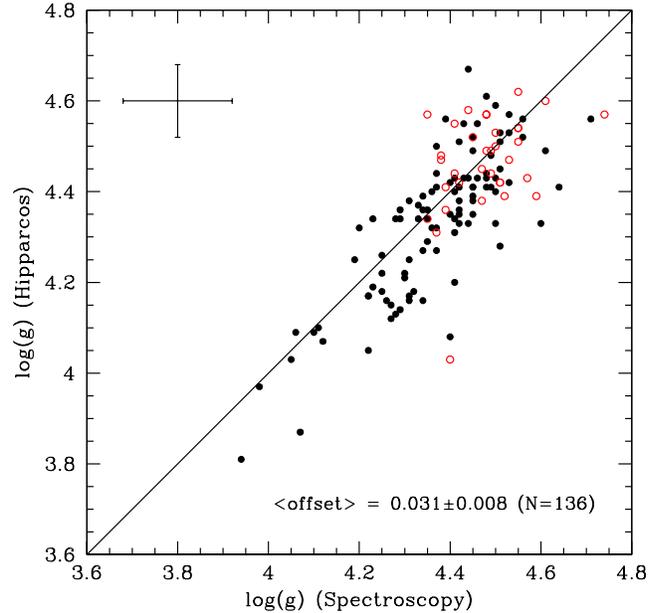}
\caption[]{Comparison of the spectroscopic and parallax based surface 
gravities of our program stars. Filled symbols represent planet-host stars,
while open symbols denote stars from our comparison sample. The error bars
represent typical relative errors in both axis. The solid line represents a 1:1 relation.}
\label{figlogghip}
\end{figure}

If we compare the spectroscopic surface gravities with the $\log{g}$ 
computed using the derived stellar masses (Fig.\,\ref{figlogghip}), the spectroscopic effective temperatures 
and the Hipparcos parallaxes (see above), the average difference we 
obtain is $\sim$0.03\,dex (spectroscopic gravities being higher), 
i.e. about 1\% -- see Fig.\,\ref{figlogghip}\footnote{Such differences are equivalent 
to errors of 7\% in the stellar mass, of 3\% in the distance, 
or of 1-2\% in the effective temperature.}. 
This difference is slightly higher for lower metallicity 
stars ([Fe/H]$<$$-$0.2\,dex), reaching 0.06\,dex, and smaller for
the remaining objects (around 0.02\,dex). The same ``gradient'' is seen 
if we analyze planet hosts and comparison sample stars separately. 
Such a difference might in fact reflect non-LTE effects
on \ion{Fe}{i} lines \citep[][]{The99}, and will be explored in more 
detail in a future paper. 

Interestingly, however, planet hosts have higher $\log{g}_{spec}-\log{g}_{hipp}$ 
(by $\sim$0.04\,dex), even though they are on average more metal-rich by $\sim$0.25\,dex. This same 
result was also noticed by \citet[][]{Law03}, and is opposite to the effect expected if
the excess metallicity observed for planet host stars were of external 
origin \citep[][]{For99}. 

An explanation for this latter inconsistency might 
be related to the fact that planet-host stars are, on average, hotter 
than our comparison sample objects by about 200\,K. Indeed, an analysis 
of our results shows a trend, of the order of 0.1\,dex/1000\,K, in the sense that 
higher T$_{\mathrm{eff}}$ stars also have higher than 
average $\log{g}_{spec}-\log{g}_{hipp}$. A comparison of our surface gravities with the 
ones of \citet[][]{Law03} and \citet[][]{ALP99} does not reveal such a clear slope, while a comparison 
of the values of the $\log{g}_{spec}$ and $\log{g}_{evol}$ derived by \citet[][]{Law03}
also shows the very same trend with effective temperature.
These results suggest that the problem might be related to the determination of
the trigonometric $\log{g}$ values (or else, all the three works have the
same bias). Sources of errors might include systematics in the bolometric corrections, perhaps related to
the fact that the calibration of \citet[][]{Flo96} used does not include a metallicity 
dependence \citep[see e.g.][]{Cay97}, or errors in the isochrones used to compute the 
stellar masses \citep[see e.g. ][]{Leb99}\footnote{Errors in the bolometric correction should not
have a significant influence on the derived stellar masses}. {The trend could also reflect
NLTE errors (although we caution that differential NLTE effects for stars with different temperature 
should not be very important for solar-type dwarfs \citep[see e.g. ][]{Ben03}), erroneous atomic 
line parameters, or problems in the stellar atmosphere models for different effective temperatures.}

Since the derived [Fe/H] values are not very sensitive to the
obtained $\log{g}$ (see above), this result 
does not affect the
derivation of accurate stellar metallicities.

\begin{figure}[t]
\psfig{width=\hsize,file=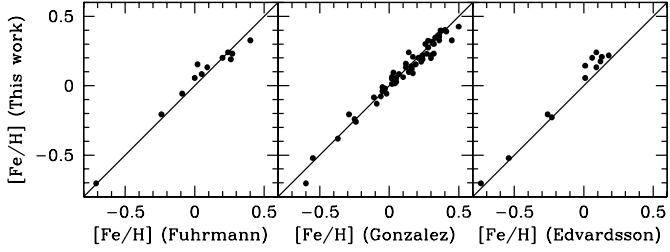}
\caption[]{Comparison of the [Fe/H] values derived in this
work with the ones obtained by other authors for the same stars. The solid line represents a
1:1 relation. See text for more details.}
\label{figcompfeh}
\end{figure}

\subsubsection{[Fe/H]}

Finally, and most importantly, we have compared our spectroscopic metallicities with the ones 
listed in all the studies mentioned above (Fig.\,\ref{figcompfeh}). The average differences found are always 
between $-$0.01 and $+$0.01\,dex, being higher only for the study \citet[][]{Edv93} 
(0.06\,dex, our results being above). In general, this difference is also not a function of the 
metallicity of the stars, i.e., within the errors it represents a uniform shift. The only marginal
trend appears when comparing our metallicities with the ones derived 
by Fuhrmann et al., in the sense that their estimates are above ours for the
more metal-rich stars, and below for the metal-poor objects.

The results we have obtained are thus perfectly compatible with other
precise published values.

\section{Other planet-host stars}

For a few planet-host stars (\object{BD$-$10\,3166}, \object{HD\,41004A}, 
\object{HD\,104985B}, and \object{GJ\,876}) we could not gather spectra and derive our own
metallicities and stellar parameters. We have thus tried to find values
of the metallicities for these stars in the literature. 
For \object{HD\,41004A}, however, there were no published spectroscopic metallicity 
estimates available, and we have
decided to obtain stellar metallicities using another technique. 

As used by several authors \citep[e.g.][]{May80,Pon97,San02},
the surface of the Cross-Correlation Function (CCF) yields precise
metallicity estimates of a star. \citet[][]{San02} (see their Appendix)
have used this method to derive a relation between [Fe/H], $B-V$, and
the surface of the CCF of the CORALIE spectrograph (hereafter $W_\mathrm{fit}$). 
This relation is now revised to take into account the slight change in the metallicity scale
introduced here, as well as metallicity estimates for new stars. The result gives:
\begin{eqnarray}
\label{eq:fehcor}
[Fe/H] & = & 2.7713+4.6826\,\log{W_\mathrm{fit}}-8.6714\,(B-V)\\
       &   & +3.8258\,(B-V)^2 \nonumber
\end{eqnarray}
a calibration valid for dwarfs with $0.52<B-V<1.09$, $1.26<W_{\mathrm{fit}}<3.14$, 
and $-0.52<[Fe/H]<0.37$. We note that the use of this relation to obtain values of 
metallicities for stars that are out of the domain of this calibration (by a small amount)
should not be of much concern, since it is expected to be a linear 
function of W$_{\mathrm{fit}}$. On the other hand, we believe it is not wise to 
extrapolate this relation for other spectral types, as the dependency 
in T$_{\mathrm{eff}}$ is much stronger and unpredictable. This calibration has an 
rms of only 0.06\,dex (N=92), similar to the typical errors of the spectroscopic 
estimates of [Fe/H]. We refer the reader to \citet[][]{San02} for more details regarding this technique.

\object{HD\,6434} was earlier reported by \citet[][]{Law03} to occupy a strange position in the HR diagram.
Curiously, when calibrating the relation expressed in Eqn.\,\ref{eq:fehcor},
\object{HD\,6434} was not included, as it was the only star falling significantly out
of the trend in the residuals of the fit. Preliminary results of a recent adaptative 
optics survey did not show the presence of any close companion to this star 
(A. Eggenberger, private communication). We do not have any
explanation for the observed discrepancy.

In Table\,\ref{tabother} we list the stellar metallicities gathered
for the stars referred above, together with their sources. For
\object{GJ\,876} alone we could not find precise metallicity estimates, as
this star is an M-dwarf.

We caution that only for \object{BD$-$10\,3166}, whose parameters were taken from the 
works of the Gonzalez team, can we be sure that 
the [Fe/H] values are in the same scale as ours. The same is true for \object{HD\,41004A},
whose [Fe/H] value was derived from Eqn.\,\ref{eq:fehcor}. For different reasons
we have chosen not to include any of these in our further analysis: \object{BD$-$10\,3166}
because it was searched for planets due to its high metal content and \object{HD\,41004A}
because its spectrum is a blend of a K and M dwarfs \citep[][]{San02}, and thus
its derived metallicity must be taken as an approximate value.

\begin{table}
\caption[]{Candidate planet-host stars for which we could not obtain a spectrum
at the time of the publication of this paper. The stellar metallicities and effective 
temperatures have been taken from various sources. For \object{HD\,41004A}, the effective 
temperature has been derived using Eqn.\,\ref{eq:teffbv} and $B-V$ taken from the 
Hipparcos catalog \citep[][]{ESA97}.}
\begin{tabular}{llll}
\hline
Star  & T$_{\mathrm{eff}}$      & \multicolumn{1}{c}{[Fe/H]}   & Source of [Fe/H]\\
      & \multicolumn{1}{l}{[K]} &                              &                 \\
\hline
\object{BD$-$10\,3166  }       & 5320 & 0.33     & \citet[][]{Gon01}     \\
\object{HD\,41004A   }       & 5085 & 0.05     & CORALIE CCF (Eqn.\,\ref{eq:fehcor}) \\
\object{HD\,104985B}$\dag$     & 4786 & $-$0.35  & \citet[][]{Sat03}    \\
\object{GJ\,876      }   & 3100-3250  & Solar  & \citet[][]{Del98}    \\
\hline
\end{tabular}
\\
$\dag$ This star is a giant
\label{tabother}
\end{table}

\section{Confirming the metal-rich nature of planet-host stars}
\label{sec:planetrich}

\begin{figure}[b]
\psfig{width=\hsize,file=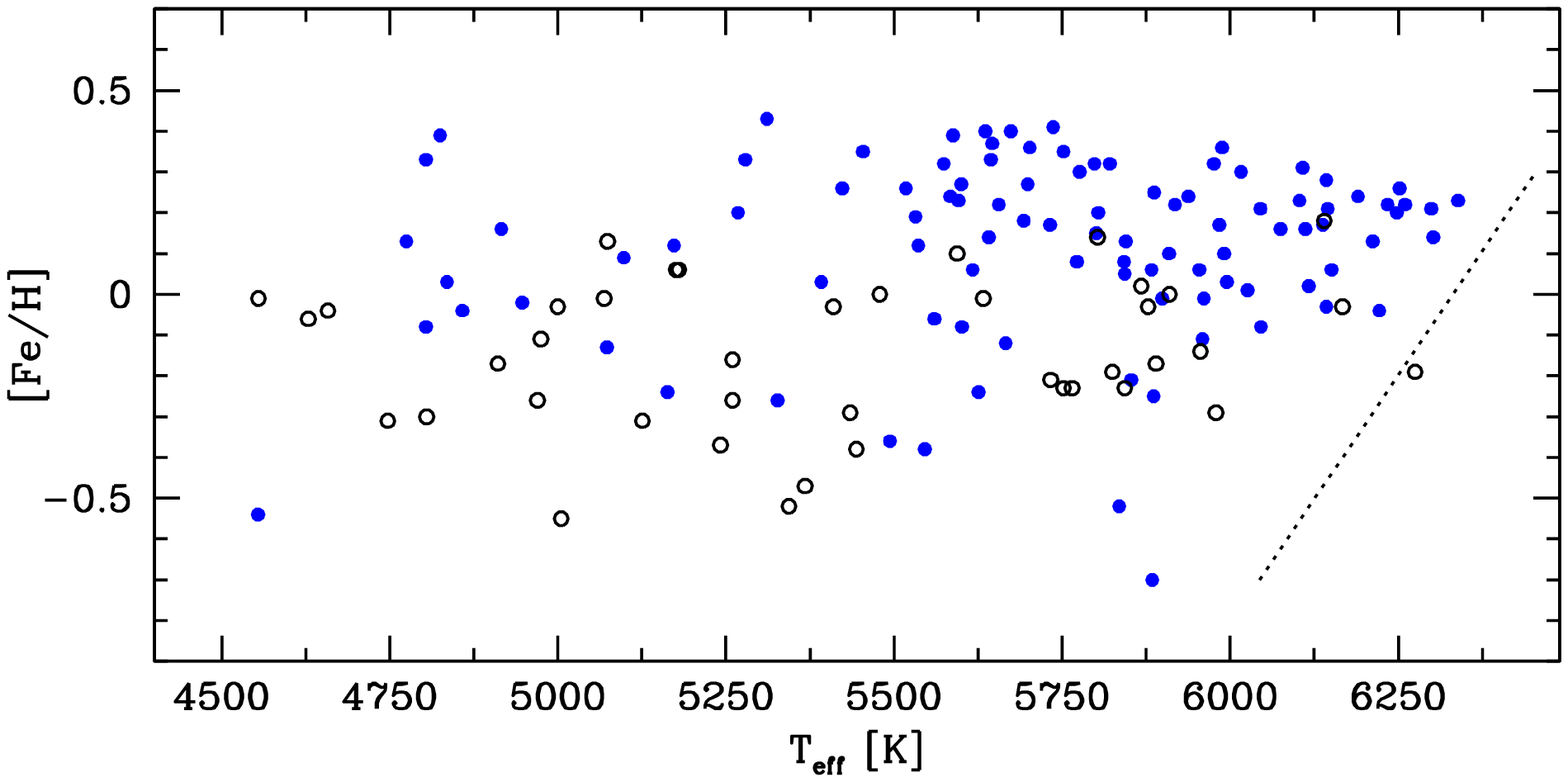}
\caption[]{Metallicity as a function of the effective temperature for
planet hosts (filled dots) and comparison sample stars (open circles).
The dotted line represents the approximate lower limit in $B-V$ of the CORALIE planet
search sample \citep[][]{Udr00}, as based on Eqn.\,\ref{eq:teffbv} (for $B-V$=constant=0.5).}
\label{figfehteff}
\end{figure}

Having gathered metallicities for almost all known exoplanet hosts,
we will now review the implications of the
available sample for the the study of the
metallicities of planet-host stars.
For an extensive discussion about the subject we point the reader to our
previous Papers\,II and III. The main difference between the current
results and the ones published in these papers are quantitative; the
qualitative results are similar.

\subsection{The global trend}

\begin{figure*}[t]
\psfig{width=\hsize,file=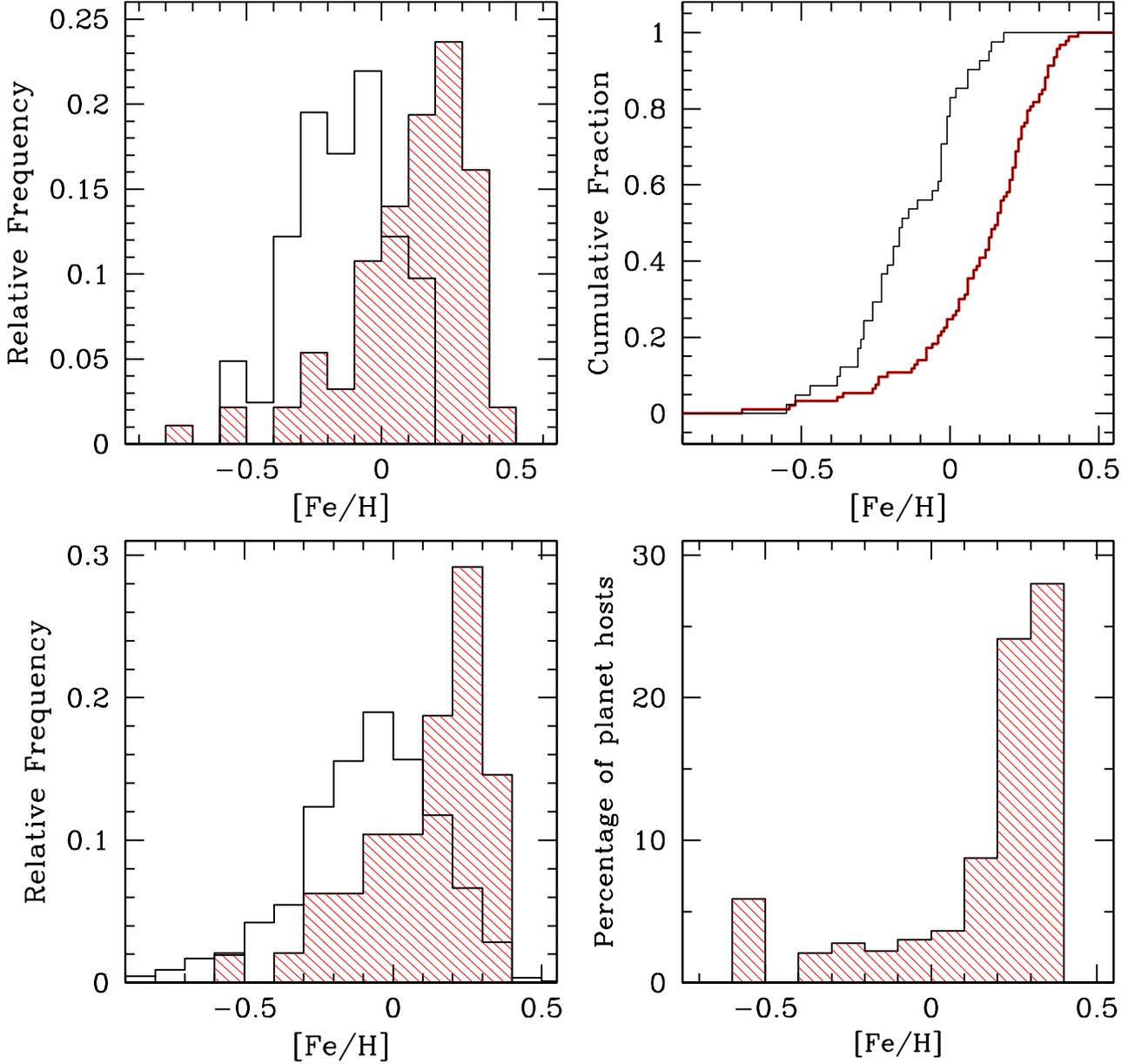}
\caption[]{{\it Upper panels}: [Fe/H] distributions for planet host stars (hashed histogram) and
for our volume-limited comparison sample of stars (open bars). The average difference between the [Fe/H]
of the two samples of $\sim$0.25\,dex. A Kolgomorov-Smirnov test
shows that the probability that the two samples are part of the same population is
of the order of 10$^{-9}$. See text for more details. {\it Lower panel, left}: [Fe/H] distributions for 
planet host stars (hashed histogram) included in the CORALIE planet-search sample, when compared with
the same distribution for all the 875 stars in the whole CORALIE program for which
we have at least 5 radial-velocity measurements (solid-line open histogram).
{\it Lower panel, right}: percentage of planet hosts found amid the stars in the CORALIE sample
as a function of stellar metallicity.}
\label{fig:histo}
\end{figure*}

In the upper panels of Fig.\,\ref{fig:histo} we present a comparison between
the metallicity distributions for our volume-limited comparison sample
of stars (Table\,\ref{tabcomp}) and for the planet-host stars with available detailed 
spectroscopic metallicities. For this latter sample, we have 
excluded those stars that were searched for planets based on their high
metallicity (we refer to Paper\,III for more details and references).
We are left with 41 stars in our comparison sample, and
with 93 planet-hosts.

A look at the two upper panels clearly shows that planet-hosts are 
considerably metal-rich compared to the comparison sample
stars by, on average, 0.25\,dex. According to a Kolmogorov-Smirnov 
test, these two samples have a probability of only 
1.6$\cdot$10$^{-9}$ of belonging to the same population. 
The results obtained with the new spectroscopic
analysis strongly confirm all the most recent results on this subject 
\citep[e.g.][]{San01,Gon01,Rei02,San03,Law03}, that show that stars with
planets are more metal-rich that average field dwarfs.

An analog of Fig.\,\ref{figfehteff} (where we plot the stellar metallicity as a function
of T$_{\mathrm{eff}}$) was used by several authors
\citep[e.g.][]{Pin01,San01,Gon01,San03} to try to decide whether the
excess metallicity observed in planet-host stars is of ``primordial origin'' (corresponding to
the metallicity of the cloud that formed the star/planet system) or of external origin
(reflecting the infall of iron-rich planetary material into the stellar
convective envelope). Although here we will not discuss this
in much detail (we refer to Paper\,III, \citet[][]{Isr03}, and \citep[][]{Gon03} for a comprehensive 
discussion), the plot of Fig.\,\ref{figfehteff}, showing that the excess metallicity found for
planet hosts is real and ``constant'' for all the T$_{\mathrm{eff}}$ regimes,
seems to support the former scenario. We should mention, however, that recent results
by \citet[][]{Vau03} suggest that this conclusion might not be straighforward; other evidence 
exist, however, supporting the promordial origin of the metallicity excess observed -- see 
Papers\,II and III.

As already noted e.g. in Papers\,II and III, a look at this figure 
also shows that the upper envelope of the planet-host metallicities is a slight decreasing
function of the stellar effective temperature. {Although not clear, this 
result may be related to the presence of NLTE effects on iron lines for
stars at different effective temperatures \citep[][]{The99}, but differential NLTE effects 
on iron lines might be relatively small in this temperature interval \citep[][]{Ben03}. }

\subsection{Planet frequency as a function of stellar metallicity}

In Fig.\,\ref{fig:histo} (lower-left panel) we compare the metallicity distribution
of the 48 planet-host stars that were found amid the dwarfs in the CORALIE
(volume-limited) planet search sample\footnote{These include the stars listed in footnote 7 of Paper\,III plus
\object{HD\,10647}, \object{HD\,65216}, \object{HD\,70642}, \object{HD\,73256}, \object{HD\,111232}, and
\object{HD\,142415}, \object{HD\,216770}.} \citep[][]{Udr00} with the [Fe/H] distribution for 
the objects in the CORALIE sample for which we have gathered at
least 5 radial-velocity measurements (solid line histogram).
The metallicities for this large sample have been obtained using Eqn.\,\ref{eq:fehcor},
and are thus in the same scale as the values obtained with our detailed
spectroscopic analysis. This sub-sample is built up of
stars for which we should have found a giant planet, at least if it
had a short period orbit.

This ``comparison'' distribution give us the opportunity to derive the
frequency of planets as a function of stellar metallicity for the stars
in the CORALIE sample. Such a result is presented in Fig.\,\ref{fig:histo} 
(lower-right panel). The figures tells us that the
probability of finding a planet is a strong function of the
stellar metallicity. About 25-30\% of the stars with [Fe/H] above 0.3 have
a planet. On the other hand, for stars with solar metallicity this percentage
is lower than 5\%. These numbers thus confirm previous qualitative results on
this matter (see Papers\,II and III, and articles by \citet[][]{Rei02} and \citet{Law03}; 
similar results were also recently presented by D. Fischer at the IAU219 symposium, 
regarding an analysis of the Lick planet survey sample). We note that in Paper\,III, 
the percentage values in Fig.\,2 are wrong by a constant factor; however, 
the results are qualitatively the same -- see also Paper\,II. 

\begin{figure}[t]
\psfig{width=\hsize,file=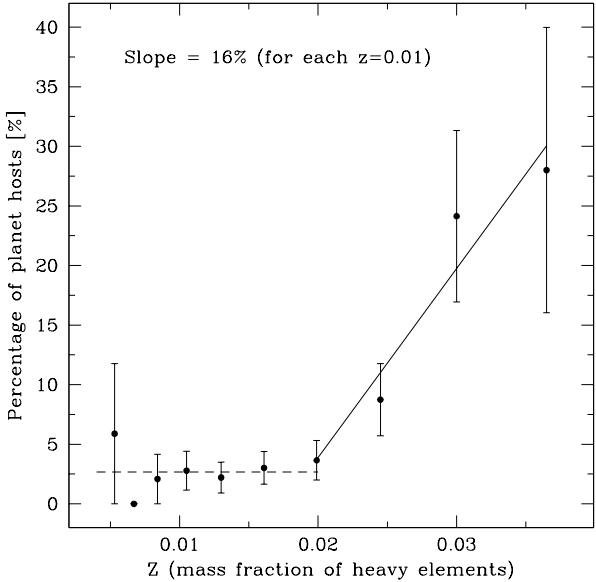}
\caption[]{Percentage of planet hosts for the plot 
in Fig.\,\ref{fig:histo} (lower-right panel, hashed histogram) as a function of the mass 
fraction of heavy elements (an increasing function of [Fe/H]). Error bars are approximate 
values based on Gaussian statistics. The plot suggests that the 
percentage is relatively constant for Z$<$0.02 (solar), increasing then linearly for 
higher Z values, with an increase of 16\% for each $\Delta$Z=0.01. }
\label{figz}
\end{figure}

The exact percentages discussed above depend mainly on the sub-sample of
stars in the CORALIE survey used to compute the frequencies. 
Current values can only be seen as lower limits, and the true numbers 
will only be known when the survey is closer to the end (although the
order of magnitude is probably the one presented here).
Only then will we also be able to provide plots
regarding e.g. stars having planets with different orbital properties 
and masses (e.g. orbital period). But present day results suggest
that there are no strong and clear correlations between stellar metallicity
and the planetary parameters (see e.g. Paper\,III).

The main interest here resides in the qualitative, rather than in
the quantitative result. The crucial conclusion 
is that more metal-rich stars seem to form planets more easily (and/or more planets?) than 
their lower-[Fe/H] counterparts. The dependence seems to be very steep, as
illustrated in Fig.\,\ref{fig:histo}. The probability of forming a planet seems to be
a strong function of the metallicity of the proto-planetary disk.
This result, valid at least for the kind of planets that are now being discovered, has enormous 
implications for the theories of planetary formation and
evolution (see Paper\,III for an extensive discussion on this subject), as
well as on studies of the frequency of planets in the galaxy \citep[e.g.][]{Lin01}.

\subsection{A flat metallicity tail?}

In Fig.\,\ref{fig:histo} (lower-right panel),
for [Fe/H]$<$0.0\,dex (Z$<$0.02), we have the impression that the corrected distributions
are rather flat (see also Fig.\,\ref{figz}). Although it is probably too early to make a definite
conclusion, if confirmed this could imply that the probability 
of forming a planet is reasonably constant for metallicities up to about the solar value, and 
only then, there is some kind of ``runaway'' 
process that considerably enhances the efficiency of planetary formation. 

In Fig.\,\ref{figz} we plot the percentage of known planets
as a function of stellar Z (the mass fraction of heavy elements). 
The plot also reflects the flatness of the distribution for metallicities
below solar (Z$<$0.02), and an increase for higher values. Curiously,
for Z$>$0.02 the percentages seem to be linearly related to Z, 
with a slope of $\sim$16\% for each $\Delta$Z=0.01. 

One possibility to explain these trends would be to consider that these reflect 
the presence of two distinct populations of exoplanets 
\citep[something already discussed in Paper III and][]{Gon03}, 
formed by different processes: one of them not dependent on the 
metallicity \citep[e.g. disk instability --][]{Bos02,Mayer02}, 
producing a constant minimum number of planets as a function of [Fe/H], together with another
very metallicity-dependent \citep[a process such as core accretion - ][]{Pol96}.
In this context, we have searched for possible differences in the
properties of the planets orbiting stars in
different metallicity regimes (eccentricity, period, masses). Nothing 
statistically significant is found \citep[see Paper~III and][]{Law03}.
In particular, no clear differences in the mass distributions 
for the planetary companions seem to exist regarding stars 
with [Fe/H]$<$0.0 and [Fe/H]$>$0.0. If indeed we were seeing two different 
populations of planets, such differences could be expected, as
disk instability processes should be able to form preferentially higher
mass planets (opposite to the core-accretion) -- \citep[see e.g. ][]{Rice03}. We note, 
however, that a slight trend in the opposite sense is found (see Paper~III),
i.e., lower metallicity stars seem to harbor preferentially lower mass planets.

A recent work by \citet[][]{Ric03} has pointed out that
giant planets might be formed in relatively metal-poor disks by the traditional core-accretion 
model (although at lower probabilities), in a timescale compatible 
with the currently accepted disk lifetimes. Indeed, core-accretion models have been
usually criticized because they predict that the formation of a giant planet could take 
longer than the estimated lifetimes of T-Tauri disks \citep[e.g.][]{Hai01}. Recent 
developments have, however, put new constraints on the
disk lifetimes that may be considerably longer than previously predicted \citep[][]{Bar03}.
Furthermore, according to \citet[][]{Ric03} the disk lifetimes might not be a problem at all.
The key to this are turbulent fluctuations in the protoplanetary disk, inducing a
``random walk'' migration, that accelerates the formation of
the giant planet \citep[][]{Ric03}. If true, this result might explain the existence of
giant planets around mildly metal-poor stars, as observed. However, the work
of Rice \& Armitage does not tell us much about the observed trends, and in particular about the
possible flatness observed in the corrected metallicity distribution for values below about solar.
Instead, it implies that disk-instability models are probably not
needed to explain the presence of giant planets around the most metal-poor stars
in our sample.

Note that the lowest metallicity bin of the plots is based
on only one planet-host, and is thus not statistically significant.

\section{Concluding remarks}   

In this paper we have derived stellar metallicities from a detailed spectroscopic
analysis of a sample of 98 stars known to be orbited by planetary mass companions,
as well as for a volume-limited sample of stars not known to host any planets.
The main results are:

\begin{itemize}

\item The obtained stellar parameters (T$_{\mathrm{eff}}$, $\log{g}$, [Fe/H], and stellar masses) 
are compatible, within the errors, with the values derived by other authors using similar or
different techniques. In particular, the derived surface gravities are only on average $\sim$0.03\,dex
different to trigonometric estimates based on Hipparcos parallaxes.

\item We confirm the previously known trends that
stars with planets are more metal-rich than average field dwarfs. The average
difference is of the order of 0.25\,dex. 

\item We confirm previous results (e.g. Papers II and III) that have shown that the
frequency of stars having planets is a strong rising function of the
stellar metallicity. About 25-30\% of dwarfs in the CORALIE planet search sample having
[Fe/H]$>$0.3 harbor a planetary companion. This number falls to $\sim$3\% 
for stars of solar metallicity. The Sun is in the tail of this distribution,
that seems to be rather flat for [Fe/H]$<$0.0 (i.e. for mass fractions of heavy elements Z$<$0.02), 
but increasing (maybe linearly) as a function of Z for higher values. Possible
implications of these results are discussed.

\end{itemize}

The main conclusions of this paper agree with previous
results that have investigated the striking role that stellar metallicity
seems to be playing in the formation of giant planets, or at least in the
formation of the kind of systems ``planet-hunters'' are finding now. However, it
is crucial that this kind of analysis is done on a continuous
basis as new planets are added to the lists. In particular,
the question of knowing whether the the Solar System is typical is particularly 
troubling, as the Sun falls in the tail of the [Fe/H] distributions of planet-host stars.

\appendix{}
\section{A calibration of T$_{\mathrm{eff}}$ as a function of $B-V$ and [Fe/H]}

\begin{figure}[t]
\psfig{width=\hsize,file=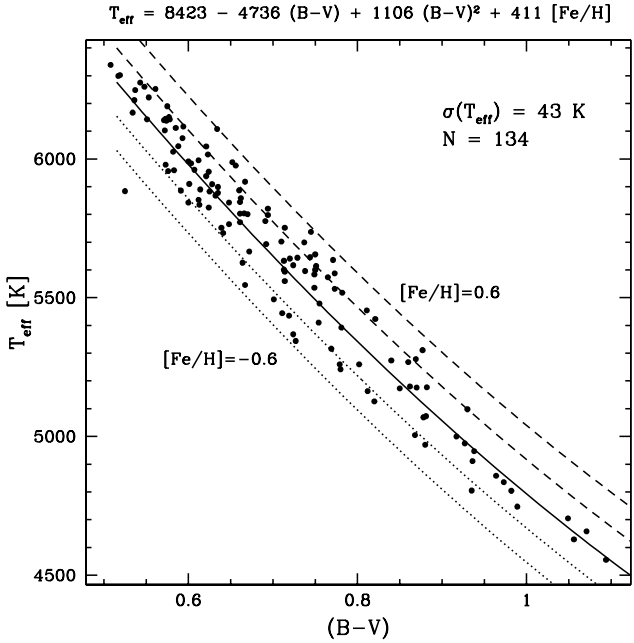}
\caption[]{Calibration of the T$_{\mathrm{eff}}$ as a function of $B-V$ and [Fe/H]. The 
5 ``fitted'' lines represent lines of constant [Fe/H] (in steps of 0.3\,dex).}
\label{figbvteff}
\end{figure}

\begin{figure}[t]
\psfig{width=\hsize,file=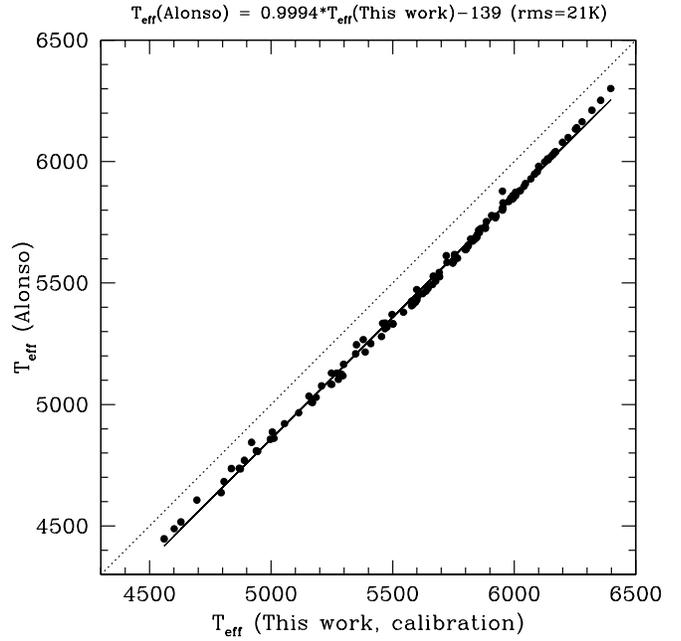}
\caption[]{Comparison between the effective temperatures derived from our calibration
and the one of \citet[][]{Alo96}. The dotted line represents a 1:1 relation
while the solid line is a linear fit to the points.}
\label{figalonso}
\end{figure}

We have used the derived spectroscopic T$_\mathrm{eff}$ and [Fe/H] as well as Hipparcos $B-V$ colors \citep[][]{ESA97} 
to derive a new calibration of the effective temperature as a function of $B-V$ and [Fe/H]. The result, also illustrated in Fig.\,\ref{figbvteff}, is:
\begin{eqnarray}
\label{eq:teffbv}
T_{\mathrm{eff}} & = & 8423 - 4736\,(B-V) + 1106\,(B-V)^2 \\
                 &   & + 411\,[Fe/H] \nonumber
\end{eqnarray}
valid for stars with $\log{g}>4.0$ in the range of $0.51<B-V<1.33$, $4495<T_{\mathrm{eff}}<6339$\,K, and $-0.70<[Fe/H]<0.43$.
The rms of the fit is only of 43\,K, illustrating the quality of the relation.
We can use this calibration to derive reliable temperatures for
our stars, whenever a detailed spectroscopic analysis is not possible,
with the guarantee that the resulting values will be in the same T$_{\mathrm{eff}}$ scale.

In Fig.\,\ref{figalonso} we compare the effective temperatures derived 
from Eqn.\,\ref{eq:teffbv} with the ones obtained from a similar calibration presented 
by \citet[][]{Alo96} for all the stars in our sample. A fit to the data gives:
\begin{eqnarray}
\label{eq:alonso}
T_{\mathrm{eff}}^{Alonso} & = & 0.9994~T_{\mathrm{eff}}^{This~work}-139 \\\nonumber
\end{eqnarray}
Except for the presence of a constant offset (reflecting different temperatures scales), the 
fit is remarkably good, having a dispersion of only 21\,K.

\begin{acknowledgements}
  We would like to thank Nami Mowlavi for the important help in 
  determining the stellar masses, David James for obtaining spectra 
  for 3 of our targets, as well as to P. Bartholdi, S. Udry, F. Pont, 
  D. Naef, and S. Jorge for fruitful discussions. We wish to thank 
  the Swiss National Science Foundation (Swiss NSF) for the continuous 
  support for this project. Support from Funda\c{c}\~ao para a Ci\^encia 
  e Tecnologia (Portugal) to N.C.S. in the form of a scholarship 
  is gratefully acknowledged.
\end{acknowledgements}


\begin{thebibliography}{}

\bibitem[Allende Prieto \& Lambert(1999)]{ALP99}
Allende Prieto, C., Lambert, D. L., 1999, A\&A, 352, 555

\bibitem[Allende Prieto et al.(1999)]{All99}
Allende Prieto, C., Garc\'{\i}a L\'opez, R., Lambert, D. L., Gustafsson, 1999, ApJ, 527, 879

\bibitem[Alonso et al.(1996)]{Alo96}
Alonso, A., Arribas, S., Martinez-Roger, C., 1996, A\&A, 313, 873


\bibitem[Bensby et al.(2003)]{Ben03}
Bensby, T., Feltzing, S., Lundstr\"om, I., 2003, A\&A, 410, 527

\bibitem[Bary et al.(2003)]{Bar03}
Bary, J.S., Weintraub, D.A., \& Kastner, J.H. 2003, ApJ, 586, 1136

\bibitem[Bergbusch \& Vanderberg(1992)]{Ber92}
Bergbusch, P.A., Vandenberg, D.A., 1992, ApJS, 81, 163

\bibitem[Bessell et al.(1998)]{Bes98}
Bessell, M.C., Castelli, F., Plez, B., 1998, A\&A, 333, 231

\bibitem[Boss(2002)]{Bos02}
Boss A., 2002, ApJ 567, L149

\bibitem[Cayrel et al.(1997)]{Cay97}
Cayrel, R., Castelli, F., Katz, D., et al., 1997, in: Proceedings of the ESA Symposium 
``Hipparcos - Venice 97'', ESA SP-402

\bibitem[Delfosse et al.(1998)]{Del98}
Delfosse, X., Forveille, T., Mayor, M., et al., 1998, A\&A, 338, L67

\bibitem[Desidera et al.(2003)]{Des03}
Desidera, S., Gratton, R.G., Endl, M., et al., 2003, A\&A, 405, 207

\bibitem[Edvardsson et al.(1993)]{Edv93}
Edvardsson, B., Andersen, J., Gustafsson, B., Lambert, D. L., Nissen, P. E., Tomkin, J., 1993, A\&A, 275, 101

\bibitem[Eggenberger et al.(2003)]{Egg03}
Eggenberger, A., Udry, S., Mayor, M., 2003, A\&A, in press

\bibitem[ESA(1997)]{ESA97}
ESA 1997, The Hipparcos and Tycho Catalogues, ESA SP-1200

\bibitem[Flower(1996)]{Flo96}
Flower, P.J., 1996, ApJ, 469, 355

\bibitem[Ford et al.(1999)]{For99}
Ford, E.B., Rasio, F.A., Sills, A., 1999, ApJ, 514, 411

\bibitem[Fuhrmann et al.(1998)]{Furetal98}
Fuhrmann, K., Pfeiffer, M.J.; Bernkopf, J., 1998, A\&A, 336, 942

\bibitem[Fuhrmann(1998)]{Fur98}
Fuhrmann, K., 1998, A\&A, 338, 161

\bibitem[Fuhrmann et al.(1997)]{Fur97}
Fuhrmann K., Pfeiffer M.J., \& Bernkopf J., 1997, A\&A 326, 1081

\bibitem[Gim\'enez(2000)]{Gim00}
Gim\'enez, A., 2000, A\&A 356, 213

\bibitem[Gonzalez et al.(2003)]{Gon03}
Gonzalez, G., 2003, Rev. Mod. Phys., 75, 101

\bibitem[Gonzalez et al.(2001)]{Gon01} 
Gonzalez G., Laws C., Tyagi S., \& Reddy B.E., 2001, AJ 121, 432

\bibitem[Gonzalez(1998)]{Gon98}
Gonzalez, G. 1998, A\&A, 334, 221

\bibitem[Gonzalez(1997)]{Gon97}
Gonzalez G., 1997, MNRAS 285, 403

\bibitem[Haisch et al.(2001)]{Hai01}
Haisch K.E. Jr., Lada E.A., \& Lada C.J. 2001, ApJ, 553, L153

\bibitem[Hatzes et al.(2003)]{Hat03}
Hatzes, A., Cochran, W., Endl, M., et al., 2003, ApJ, submitted

\bibitem[Heiter \& Luck(2003)]{Hei03}
Heiter, U., Luck, R.E., 2003, AJ, 126, 2015

\bibitem[Israelian et al.(2003)]{Isr03}
Israelian, G., Santos, N.C., Mayor, M., Rebolo, R., 2003, A\&A, 405, 753 
 
\bibitem[Jorissen et al.(2001)]{Jor01}
Jorissen, A., Mayor, M., \& Udry, S., 2001, A\&A 379, 992

\bibitem[Jones et al.(2002)]{Jon02}
Jones, H.R.A., Butler, R.P., Tinney, C.G., 2002, MNRAS, 333, 871

\bibitem[Kurucz(1993)]{Kur93} 
Kurucz, R. L., 1993, CD-ROMs, ATLAS9 Stellar
Atmospheres Programs and 2~${\rm km}~{\rm s}^{-1}$ Grid
(Cambridge: Smithsonian Astrophys. Obs.)

\bibitem[Kurucz et al.(1984)]{Kur84} 
Kurucz, R. L., Furenlid, I., Brault, J., Testerman, L., 1984, Solar Flux Atlas 
from 296 to 1300 nm, NOAO Atlas No.\,1

\bibitem[Laws et al.(2003)]{Law03}
Laws, C., Gonzalez, G., Walker, K.M., et al., 2003, AJ 125, 2664

\bibitem[Lebreton et al.(1999)]{Leb99}
Lebreton, Y., Perrin, M.-N., Cayrel, R., Baglin, A., Fernandes, J., 1999, A\&A, 350, 587

\bibitem[Lineweaver(2001)]{Lin01}
Lineweaver, C.H., 2001, Icarus, 151, 307

\bibitem[Lutz \& Kelker(1973)]{Lut73}
Lutz, T.E., Kelker, D.H., 1973, PASP, 85, 573

\bibitem[Martell \& Laughlin(2002)]{Martell02}
Martell, S., Laughlin, G., 2002, ApJ 577, L45

\bibitem[Mayer et al.(2002)]{Mayer02} 
Mayer, L., Quinn, T., Wadsley, J., Stadel, J., 2002, Science, 298, 1756

\bibitem[Mayor(2003)]{May03b} 
Mayor, M., 2003, in Extrasolar Planets: Today and Tomorrow, ed. C. Terquem, A. 
Lecavelier des Etangs, \& J.-P. Beaulieu, XIXth IAP Colloqium, in press

\bibitem[Mayor et al.(2003)]{May03} 
Mayor, M., Udry, S., Naef, D., et al. 2003, A\&A, in press


\bibitem[Mayor(1980)]{May80}
Mayor, M., 1980, A\&A, 87, L1

\bibitem[Murray \& Chaboyer(2002)]{Mur02}
Murray, N., Chaboyer, B., 2002, ApJ 566, 442

\bibitem[Nissen et al.(1997)]{Nis97}
Nissen, P. E., Hoeg, E., Schuster, W. J., 1997, in: Proceedings of the ESA Symposium 
``Hipparcos - Venice 97'', ESA SP-402

\bibitem[Pinsonneault et al.(2001)]{Pin01} 
Pinsonneault, M.H., DePoy, D.L., \& Coffee, M. 2001, ApJ 556, L59

\bibitem[Pollack et al.(1996)]{Pol96}
Pollack, J.B., Hubickyj, O., Bodenheimer, P., et al., 1996, Icarus 124, 62

\bibitem[Pont(1997)]{Pon97}
Pont, F., Ph.D. Thesis, Geneva University

\bibitem[Press et al.(1992)]{Numerical}
Press, W.H., Teukolsky, S.A., Vetterling, W.T., \& Flannery, B.P. 1992, In: ``Numerical Recipes in Fortran 77'', Volume 1, Cambridge University Press, Second Edition


\bibitem[Reid(2002)]{Rei02}
Reid, I.N., 2002, PASP 114, 306

\bibitem[Ribas et al.(2003)]{Rib03}
Ribas, I., Solano, E., Masana, E., \& Gimenez, A., 2003, A\&A, in press (astro-ph/0310456)

\bibitem[Rice et al.(2003)]{Rice03}
Rice, W.K.M., Armitage, P., Bonnel, I.A., Bate, M.R., Jeffers, M.M., \& Vine, S.G., 2003, MNRAS, in press (astro-ph/0310679)

\bibitem[Rice \& Armitage(2003)]{Ric03}
Rice, W.K.M., \& Armitage, P., 2003, ApJ, 598, L55

\bibitem[Santos et al.(2003a)]{San03} 
Santos, N.C., Israelian, G., Mayor, M., Rebolo, R., \& Udry, S., 2003a, A\&A  398, 363 (Paper\,III)

\bibitem[Santos et al.(2003b)]{San03b} 
Santos, N.C., Mayor, M., Udry, S., et al., 2003b, in: Proceedings of the IAU symposium 219 ``Stars as Suns: Activity, Evolution, \& Planets'', in press

\bibitem[Santos et al.(2002)]{San02}
Santos, N.C., Mayor, M., Naef, D., et al., 2002, A\&A, 392, 215

\bibitem[Santos et al.(2001)]{San01} 
Santos, N.C., Israelian, G., \& Mayor, M., 2001a, A\&A, 373, 1019 (Paper\,II)

\bibitem[Santos et al.(2000)]{San00} 
Santos N.C., Israelian G., \& Mayor M., 2000, A\&A 363, 228 (Paper\,I)

\bibitem[Sato et al.(2003)]{Sat03}
Sato, B., Ando, H., Kambe, E., et al., 2003, ApJ, 597, L157

\bibitem[Schaerer et al.(1993)]{Sch93} 
Schaerer, D., Charbonnel, C., Meynet, G., et al., 1993, A\&AS 102, 339

\bibitem[Schaerer et al.(1992)]{Schae92} 
Schaerer, D., Meynet, G., Maeder, A., et al., 1992, A\&AS 98, 523

\bibitem[Schaller et al.(1992)]{Sch92} 
Schaller, G., Schaerer, D., Meynet, G., \& Maeder, A., 1992, A\&AS 96, 269


\bibitem[Smith(2003)]{Smi03}
Smith, H.Jr., 2003, MNRAS, 338, 891

\bibitem[Sneden(1973)]{Sne73} 
Sneden, C., 1973, Ph.D. thesis, University of Texas

\bibitem[Th\'evenin \& Idiart(1999)]{The99}
Th\'evenin, F., \& Idiart, T.P., 1999, ApJ 521, 753

\bibitem[Udry et al.(2003)]{Udr03}
Udry, S., Mayor, M., Santos, N.C., 2003, A\&A, 407, 369

\bibitem[Udry et al.(2000)]{Udr00}
Udry, S., Mayor, M., Naef., D., et al., 2000, A\&A, 356, 599

\bibitem[Vauclair(2003)]{Vau03}
Vauclair, S., 2003, ApJ, submitted (astro-ph/0309790)

\bibitem[Zucker \& Mazeh(2002)]{Zuc02}
Zucker, S., \& Mazeh, T., 2002, ApJ 568, L113

\bibitem[Zucker et al.(2001)]{Zuc01}
Zucker, S., Naef, D., Latham, D., et al., 2001, ApJ., 568, 363

\end{thebibliography}
\end{document}